\documentclass[10pt,twocolumn,letterpaper]{article}

\usepackage{cvpr}              %
\usepackage{todonotes}
\usepackage{capt-of}

\usepackage{overpic}

\newcommand{\fsdf}[0]{f_{\mathrm{sdf}}}
\newcommand{\fdeform}[0]{f_{\mathrm{deform}}}
\newcommand{\Rthree}[0]{\mathbb{R}^3}

\definecolor{cvprblue}{rgb}{0.21,0.49,0.74}
\usepackage[pagebackref,breaklinks,colorlinks,citecolor=cvprblue]{hyperref}

\def\paperID{100} %
\def\confName{3DV\xspace}
\def\confYear{2026\xspace}

\title{MoAngelo: Motion-Aware Neural Surface Reconstruction for Dynamic Scenes}

\author{Mohamed Ebbed\\
\small{University of Bonn} \\
\small{Lamarr Institute for Machine Learning} \\
\small{and Artificial Intelligence} \\
{\tt\small mebbed@uni-bonn.de}
\and
Zorah Lähner\\
\small{University of Bonn}\\
\small{Lamarr Institute for Machine Learning} \\
\small{and Artificial Intelligence} \\
{\tt\small laehner@uni-bonn.de}
}

\begin{document}
\twocolumn[{%
\renewcommand\twocolumn[1][]{#1}%

  \maketitle
  \begin{center}
    \includegraphics[width=\linewidth]{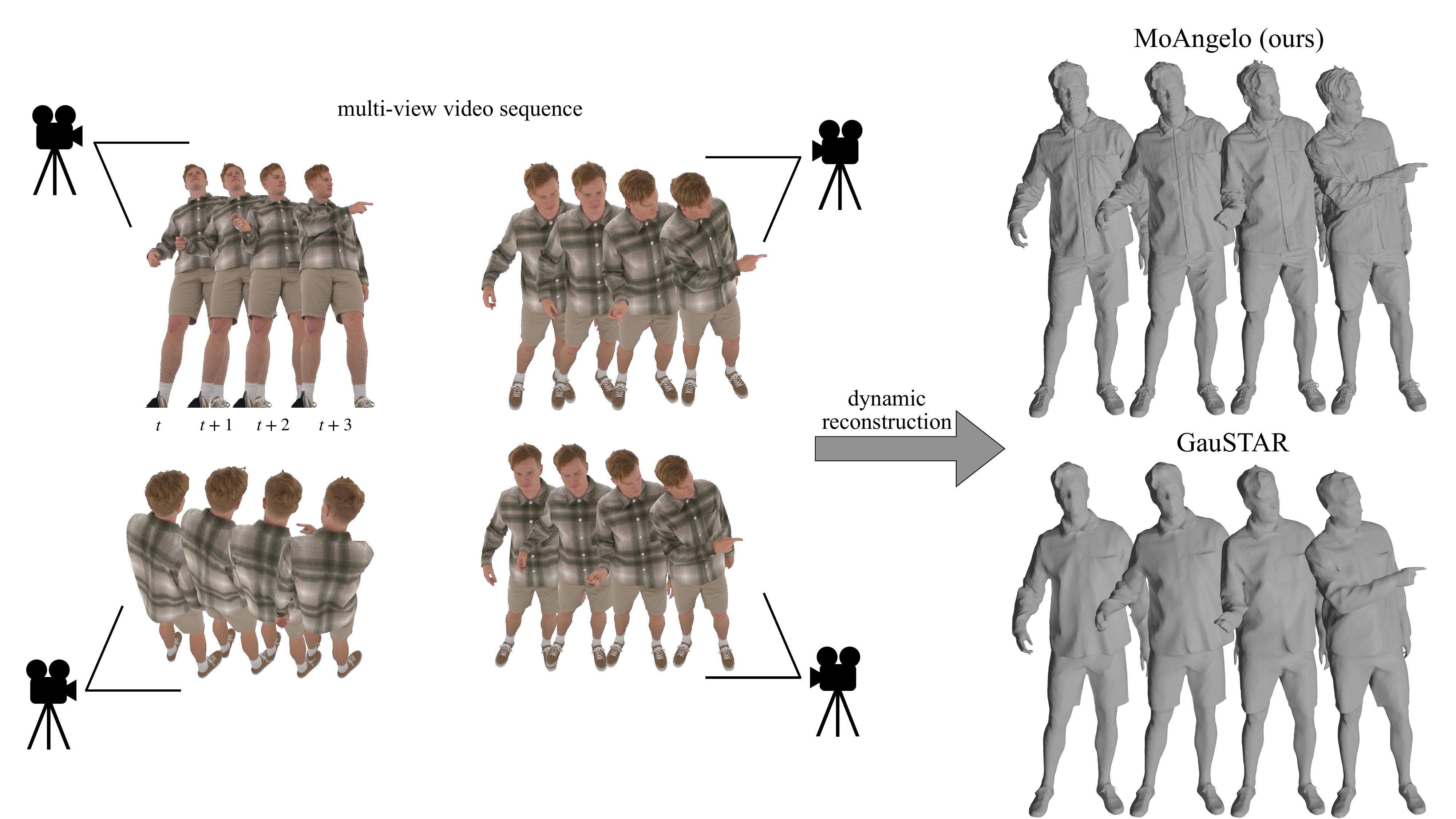}
    \captionof{figure}{We present MoAngelo, a dynamic multi-view reconstruction method, that can produce more detailed geometry than competitors and avoids smoothing out details. Our method is based on a static template reconstruction using NeuralAngelo~\cite{li2023neuralangelo}, however, in contrast previous approaches, our template is flexible and refined during the optimization of the deformation fields which leads to better accuracy.}
    \label{fig:teaser}
  \end{center}
}]
\begin{abstract}
Dynamic scene reconstruction from multi-view videos remains a fundamental challenge in computer vision.
While recent neural surface reconstruction methods have achieved remarkable results in static 3D reconstruction, extending these approaches with comparable quality for dynamic scenes introduces significant computational and representational challenges. 
Existing dynamic methods focus on novel-view synthesis, therefore, their extracted meshes tend to be noisy. 
Even approaches aiming for geometric fidelity often result in too smooth meshes due to the ill-posedness of the problem.
We present a novel framework for highly detailed dynamic reconstruction that extends the static 3D reconstruction method NeuralAngelo to work in dynamic settings.
To that end, we start with a high-quality template scene reconstruction from the initial frame using NeuralAngelo, and then jointly optimize deformation fields that track the template and refine it based on the temporal sequence.
This flexible template allows updating the geometry to include changes that cannot be modeled with the deformation field, for instance occluded parts or the changes in the topology.
We show superior reconstruction accuracy in comparison to previous state-of-the-art methods on the ActorsHQ dataset.
\end{abstract}
    
\section{Introduction}

Dynamic reconstruction, the task of recovering the dynamic geometry of a scene captured from multi-view videos, remains a challenging problem with many applications in virtual and augmented reality, scene understanding and dynamic asset creation. 
Recent advances in reconstruction and novel view synthesis have shown impressive results for static scenes~\cite{kerbl3Dgaussians,li2023neuralangelo} but moving objects increase the complexity and computational demands of the setup substantially. 
The core difficulty in dynamic reconstruction stems from the ill-posedness of the problem: recovering time-varying 3D structures from sparse viewpoints requires priors and regularization to disambiguate between changes caused by motion and viewpoint variations.
Early methods relied on pre-defined templates and structured motion models. 
While these approaches work well on controlled datasets, they are limited by their reliance on explicit surface representations and hand-crafted motion priors, making them struggle with complex deformations and topological changes. 

Neural Radiance Fields (NeRF)~\cite{mildenhall2020nerf, barron2021mipnerfmultiscalerepresentationantialiasing, mueller2022instant} have revolutionized static 3D scene representation by demonstrating that neural fields can capture view-dependent appearance effects. 
Despite the impressive novel-view synthesis results, the volume rendering approach neglects the representation of geometry, which makes it difficult to extract 3D meshes. %
As a solution to this, Neural SDFs~\cite{li2023neuralangelo, wang2023neuslearningneuralimplicit, yariv2021volume} represent the geometry as a neural signed distance function (SDF) instead of a volume density, enabling the extraction of high-quality surfaces by applying marching cubes~\cite{lorensen1987marchingcubes}.
However, these methods are limited to static scenes and it remains challenging to extend them to dynamic scenes with the same quality of results.

Due to the overall success of neural radiance fields, several methods have extended them for dynamic settings by adding a temporal dimension or a deformation field~\cite{pumarola2020dnerfneuralradiancefields, park2021nerfies, tretschk2024scenerflow, isik2023humanrf}.
However, the same difficulties with extracting meshes from volume densities remain.  
Current trends in dynamic scene representation rely heavily on deformable 3D Gaussian splatting (3DGS)~\cite{yu2024gaussian,4dgaussians} which are spare and, thus, more efficient to optimize but also target mainly novel-view synthesis.
Some methods represent the moving geometry as a neural SDF~\cite{shao2023tensor4d} or track 3DGS together with a mesh~\cite{zheng2025gaustar}.
In these cases, the extracted geometry lacks fine geometric details and the reconstruction quality is not comparable with the quality of static 3D reconstruction methods. 
The reasons range from the template being implicit only which averages the geometry through the sequence making it overly smooth~\cite{shao2023tensor4d}, to limited mesh resolution for optimization reasons~\cite{zheng2025gaustar}.%
In summary, dynamic scene representation methods are achieving impressive novel-view synthesis results but lack the capability to extract high fidelity geometry.

In this work, we propose a novel framework to do high-fidelity dynamic neural surface reconstruction. 
Our idea is based on taking NeuralAngelo~\cite{li2023neuralangelo}, the current state-of-the-art in static reconstruction based on neural SDFs with highly detailed reconstructions, on the first frame, and then extending it to iteratively learn neural deformation fields that track the movement of the template through time while jointly refining its the geometry. 

\paragraph{Contributions. } Our main contributions are as follows:
\begin{itemize}
    \item A novel framework for dynamic reconstruction that jointly deforms the reconstructed template of an initial frame while tracking the movements of it and refining its geometry at the same time.
    \item Highly detailed dynamic reconstructions from multi-view videos that preserve surface details even in longer sequences with large movement.
    \item Our experimental results show that we outperform previous methods both qualitatively and quantitatively by a large margin. 
\end{itemize}
We will release our code upon acceptance.

\section{Related Work}

We focus this section on directly related work using RGB data but a more general survey including depth information can be found in~\cite{zollhoefer2018starrgbd}.

\subsection{Static 3D Reconstruction}

Classical multi-view reconstruction algorithms used to rely on feature extraction and matching~\cite{low1999sift,furukuwa2015mvs} or photometric stereo by using lighting variations to recover surface normal information~\cite{Haefner_2019_photometric,ackermann2015photometric}.
These methods typically produces explicit surface representations such as point clouds or meshes, which provide a direct geometric interpretation but struggle with fine details.
Recently, 3D Gaussian splatting~\cite{kerbl3Dgaussians} has been established a powerful explicit representation that combines the benefit of point-based methods with differentiable rendering. 
Even though this representation was developed for novel view synthesis, countless extensions have been proposed to improve geometric fidelity and the quality of surface extractions~\cite{yu2024gaussian,chen2024pgsr,guedon2023sugar}.

\paragraph*{Neural Field-based Methods}
A completely different approach is taken by neural field-based methods which implicitly define the scene properties like geometry or radiance by functions encoded in a neural network. 
This direction became increasingly popular after the introduction of Neural Radiance Fields (NeRF)~\cite{mildenhall2020nerf} for novel view synthesis, but implicit representations have been widely used for reconstruction before due to the flexibility in optimization~\cite{calakli2011ssd,mescheder2019occupancy}.
The quickly growing field addressed existing challenges with implicit representations through various architectural and algorithmic improvements: Instant-NGP~\cite{mueller2022instant} drastically accelerated training and inference through multi-resolution hashgrids.
In the direction of pure geometry representation, NeuS~\cite{wang2023neuslearningneuralimplicit} replace the NeRF density-based formulation with a pure signed distance function (SDF) to enable straight forward surface extract and \cite{yariv2021volume} closed the gap between volume and surface rendering by learning the relationship between a density and signed distance function. 
The highest level of geometric detail has been achieved by combining hashgrids with SDFs which was used in NeuS2~\cite{wang2023neus2}, MonoSDF \cite{Yu2022MonoSDF}, and NeuralAngelo~\cite{li2023neuralangelo}.
Our dynamic reconstruction is based on extending Neuralangelo to work for dynamic scenes by initially reconstructing a high-quality template mesh, then jointly track its movement using deformation fields and furtherly tuning the geometry of the template mesh.

\subsection{Dynamic 3D Reconstruction}

Dynamic reconstruction adds an additional time dimension to the problem which increases the computational complexity significantly. 
Early methods relied on pre-defined templates and structured motion models. For example, non-rigid structure-from-motion extended classical bundle adjustment to handle deformable objects by assuming low-rank deformation~\cite{bregler2000nonrigid}.
Multi-view stereo techniques, on the other hand, were able to incorporate temporal consistency through optical flow and surface tracking~\cite{furukuwa2015mvs,Mustafa2016temporally}.

The success of neural fields in static reconstruction naturally led to extensions in the dynamic setting. 
Nerfies~\cite{park2021nerfies} and D-NeRF \cite{pumarola2020dnerfneuralradiancefields} decomposed the problem into optimizing for a canonical NeRF that represents the canonical scene properties, and a deformation field representing the non-rigid deformations of the canonical scene. 
We follow a similar idea, instead of implicitly optimizing for the canonical scene, we explicitly define it as the initial frame of the scene and initially reconstruct it and furtherly tune its geometry through time. We also iteratively optimize for a deformation field that tracks the movement of that template mesh. SceNeRFlow \cite{tretschk2024scenerflow} also defines the template scene as the first frame and iteratively optimize for a deformation field to track it, howerver their approach is mainly focused on novel-view synthesis because the geometry is represented as a volume density, therefore it is not straightforward to extract dynamic meshes from their representation. On the other hand, our geometry is represented as a neural SDF, making it easy to extract high-resolutional meshes using marching cubes. Additionally, the current trend is to use dynamic 3D gaussians that deform through time \cite{4dgaussians}, however these approaches suffer from the same problem that they are only focused on novel-view synthesis, and suffer from the same problems that face gaussian splatting methods \cite{yu2024gaussian} when extracting a 3D mesh from them, thus all the current methods are limited only to novel-view synthesis.
We refer the reader to \cite{tretschk2023star3dnonrigid} for an in-depth survey of recent non-rigid reconstruction methods.  

Most related to our work, and thus used as competitors in the experiments section, are the following works:
HumanRF~\cite{isik2023humanrf} is able to produce high resolution 4D reconstructions by generating a 4D volume density, but the volume density leads to noisy mesh extractions.
The method of Tensor4D~\cite{shao2023tensor4d} implicitly optimizes for a template scene and a deformation field represented by 4D feature grid followed by a shallow MLP. The feature grid is decomposed into low-rank tensors using tensor decomposition methods \cite{chen2022tensorftensorialradiancefields}. The template scene is implicitly optimized for, therefore the resulted template scene is very smooth, leading to reconstructed meshes that lack the fine geometric details.
Not limited to neural fields, GauSTAR~\cite{zheng2025gaustar} combines dynamic 3D Gaussian splatting with a template reconstructed from HumanRF and deformed based on the movement of the Gaussians, however the resulted meshes is not high-resolutional, so it doesn't capture the fine geometric details.

\section{Problem Setup and Notation} \label{sec:setup}

This section will introduce the problem setup, representations and notation (\cref{tab:notation}) we used. 

\begin{table}[h]
    \centering
    \begin{tabular}{ll}
        \hline
        \textbf{Symbol} & \textbf{Description} \\
        \hline
        $t$ & time step variable, frame index \\
        $\fsdf$ & signed distance function of geometry, \\
        & also called template and canonical frame \\
        $f_{rgb}$ & the radiance field of the template frame \\
        $x_{i,t}$ & 3D coordinate of point $i$ in frame $t$ \\
        & indices are dropped when irrelevant\\
        $\hat x_{i,t}$ & $x_{i,t}$ deformed into canonical frame \\
        $s_i, d_i$ & SDF value and geometric features of $x_i$ \\
        $f^t_{deform}$ & deformation field for frame $t$ \\
        \hline
    \end{tabular}
    \caption{Overview of most important notation.}
    \label{tab:notation}
\end{table}

\subsection{Objective} \label{sub:setup:problem}
Given a sequence of multi-view RGB images of a scene captured from \( N_C \) cameras, each consisting of \( N_F \) frames, our objective is to reconstruct \( N_F \) high-quality triangular meshes, where each mesh \( M_i \) represents the detailed geometry of the scene at time step \( i \). Optionally, segmentation masks may be provided for each frame to focus only on the object of interest.

\subsection{Geometry Representation} \label{sub:setup:geometry}
The scene geometry is represented as signed distance function (SDF) $\fsdf: \mathbb{R}^3 \to \mathbb{R}$ in a multi-resolution hash grid~\cite{mueller2022instant} followed by a multi-layer perceptron (MLP) with a single hidden layer of 64 neurons. 
The SDF returns the distance to the closest surface, with the sign indicating whether the point lies inside or outside the surface. A triangular mesh can be extracted from the SDF using the marching cubes algorithm~\cite{mcubes}.
The hash grid comprises 16 resolution levels, ranging from \( 32^3 \) to \( 2024^3 \). Each level outputs a 4-dimensional feature vector, which are concatenated to form a total feature vector of dimension 64. 
This feature vector is then concatenated with the input 3D position \( x \), and the resulting vector is passed to the MLP to predict both the SDF value \( s \) at that point and an additional feature vector \( d \). 
The surface normal at \( x \) is computed as the gradient of the SDF with respect to the input position, i.e., \( n = \nabla_x f_{\mathrm{sdf}}(x) \). This gradient can be efficiently calculated using PyTorch's automatic differentiation~\cite{paszke2019pytorchimperativestylehighperformance}.
At the beginning, $f_{\mathrm{sdf}}$ is initialized by running NeuralAngelo~\cite{li2023neuralangelo} on the first frame but tracked and gradually refined during our optimization.

\subsection{Appearance Representation}
The appearance of the scene is represented using a radiance field \( f_{\mathrm{rgb}}(x, v, n, d) \), which maps the 3D position \( x \), viewing direction \( v \), surface normal \( n \), and the SDF feature vector \( d \) to a color value \( c \).

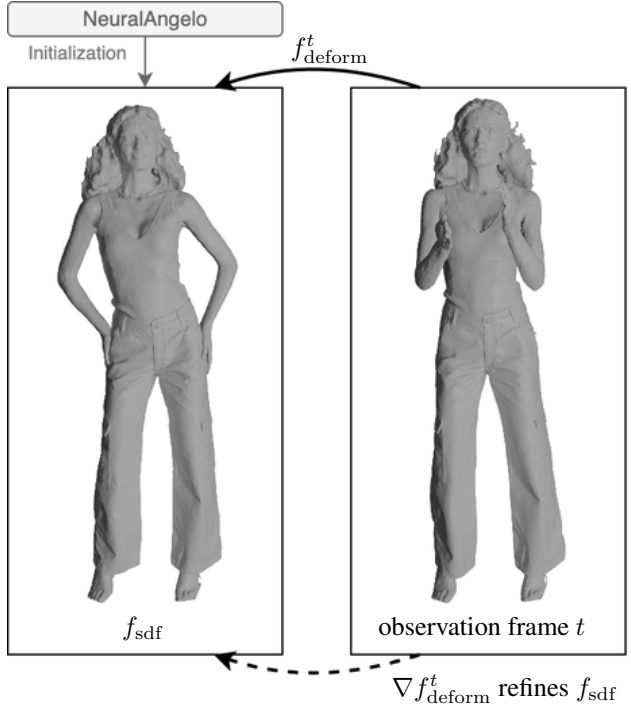
\begin{figure}
    \centering
    \begin{overpic}[width=\linewidth]{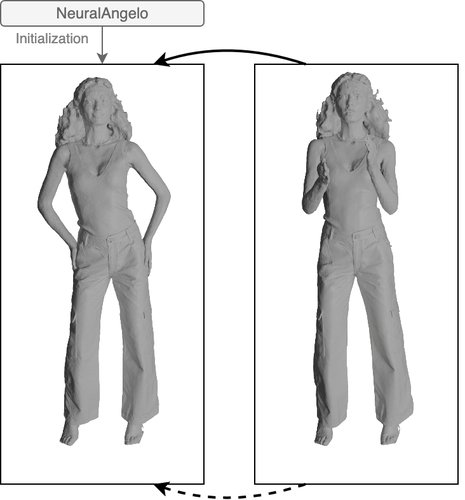}
    \put(17,6){$\fsdf$}
    \put(55,6){observation frame $t$}
    \put(42,92){$\fdeform^t$}
    \put(57,-3){$\nabla\fdeform^t$ refines $\fsdf$}
    \end{overpic}
    \vspace{0.1cm} 
    \caption{Overview of the joint optimization of the template $\fsdf$ and the deformation to each observation frame $\fdeform^t$. The movement for frame $t$ is stored in a deformation field $\fdeform^t$ and optimized based on the current state of the template $\fsdf$. However, the gradient information of this update is propagated to $\fsdf$ which is refined in turn. }
    \label{fig:framework}
\end{figure}

\section{Methodology}

We propose a novel framework for dynamic reconstruction based on explicit modeling of the template reconstructed from the canonical frame, and joint optimization of motion and template for the rest of the sequence. 
Our method starts with a static reconstruction of the first (or canonical) frame using NeuralAngelo~\cite{li2023neuralangelo} (see \cref{sub:method:template}).
In contrast to previous work, we use this reconstruction as an explicit representation that is regularized and refined during the optimization of the motion (see \cref{sub:method:representation}) morphing it into later frames, see \cref{sub:method:tracking}. 
This combination results in a high fidelity reconstruction of fine details.

\subsection{Motion Representation} \label{sub:method:representation}

We represent the scene motion as a collection of deformation fields, each mapping points from the coordinates of the current time step $t$ (observation frame) into the canonical frame. 
The geometry at time $t$ can then be computed as
\begin{equation}
    \fsdf^t(x) = \fsdf(\fdeform^t(x)),
\end{equation}
where $x \in \Rthree$, $\fdeform: \Rthree \to \Rthree$, and $\fsdf: \Rthree \to \mathbb{R}$ is the signed distance representation of the template. 
Each $\fdeform^t$ consists of a multi-resolution hash grid with the same number of levels, feature dimensions, and resolutions as the hash grid as $\fsdf$. The hash grid is followed by single-layer MLP with 64 neurons.
(see \cref{sub:setup:geometry}).
Specifically, the deformation fields predict an \( \mathfrak{so}(3) \) transformation from the observation frame to canonical frame, then the transformation is converted to a rotation matrix $R_i$ and a translation vector $T_i$ using exponential mapping \cite{Lynch_Park_2017}. We apply the transformation to $x_{i,t}$ as follows:

\begin{align}
R_{i}, T_i &= f^t_{deform}(x_{i,t}) \\
\hat{x_i} &= R_ix_{i,t} + T_i
\end{align}

In contrast to previous work~\cite{pumarola2020dnerfneuralradiancefields, shao2023tensor4d}, we model and deform the template $\fsdf$ explicitly, and can optimize its geometry and appearance to high fidelity and without oversmoothing.
However, this also requires explicit tracking of the template in all frames which we describe in the next section.

\subsection{Template Mesh Tracking} \label{sub:method:tracking}

Each deformation field $\fdeform^t$ needs to map the geometry of the corresponding frame $t$ exactly to the right position of the template geometry $\fsdf$ which will be optimized through volume rendering.

\paragraph{Volume Rendering. } 
For each frame $t$, we need to render the 3D neural fields into 2D to compare them to the input images. 
To that end, we shoot $N_r$ rays at random pixels and 80 many 3D points $x_i$ along these rays. 
The signed distance value $s_i$ and feature vector $d_i$ for $x_i$ can be computed in the canonical frame by taking $s_i, d_i = f_{sdf}(\hat{x_i})$ and the normal vector as $n_i = \nabla_x f_{\mathrm{sdf}}(\hat{x}_i)$ . 
$R_i$ is applied to the normals to move them from canonical space back to observation frame.
Finally, the color $c_i$ at $x_i$ is computed as $c_i = f_{rgb}(\hat{x}_i, v_i, d_i, n_i)$ where $v_i$ is the viewing direction of the ray. 

To render the correct RGB value $C_r$ for the pixel, we need to integrate all points on the ray shooting through it. 
\begin{equation}
C_r = \sum_{i=1}^{N_s} T_i \alpha_i \cdot c_i
\end{equation}
where $\alpha_i$ is the opacity of $x_i$ and $T_i = \prod_{j=1}^{i-1} (1 - \alpha_j)$ the transmittance which represents the probability a ray will reach the sample without being occluded, the standard way to compute the weighting \cite{wang2023neuslearningneuralimplicit, li2023neuralangelo}.

\[
\alpha_i = \max \left( \frac{\Phi_s(s_i) - \Phi_s(s_{i+1})}{\Phi_s(s_i)},\ 0 \right),
\]

Where $\Phi_s$ is the derivative of the sigmoid function, known as the logistic density distribution.  

\subsubsection{Optimization} \label{sub:method:optimization}
The deformation field $\fdeform^t$ for each frame $t$ is optimized separately but initialized with $\fdeform^{t-1}$, or identity for $t=0$.
The total loss function is 
\begin{equation}
    \mathcal{L}_{total} = L_{render} + \lambda_{mask}\mathcal{L}_{mask} +  \lambda_{eik}\mathcal{L}_{eik},
\end{equation}
definitions of each term follow below.
The loss function is very similar to a static single frame reconstruction using NeuralAngelo~\cite{li2023neuralangelo} (see \cref{sub:method:template}), In each observation frame $t$, we optimize for the current $f^t_{deform}$.
Not all geometric details were necessarily visible during initialization, due to (self-)occlusions or topology changes, therefore we refine the template scene to handle these changes by allowing the gradient signal back to $f_{sdf}$ and $f_{rgb}$. The joint optimization of the template and each deformation field enables a refinement of the geometry and appearance which leads to much more accurate and temporal consistent reconstructions than a static template or implicit template frame can.

While reconstructing the template, we run the optimization for $250k$ steps with AdamW optimizer \cite{adamw}, then for the following time steps we run it for $25k$ steps for each time step $t$. We also optimize the hash grid of each deformation field in a coarse-to-fine manner to prevent overfitting. Initially, we start the optimization with 4 active levels, and every $1000$ steps we activate an additional level until all 16 levels are active.

\paragraph*{Rendering Loss. }
To optimize the template scene, we sample a minibatch of $N_r$ pixels from a random image with ground truth RGB value of $\hat{C}_r$ and compute the L1 re-rendering loss between the rendered RGB colors of the shooted rays to these pixels and their ground truth values as follows:

\begin{equation}
\mathcal{L}_{\text{render}} = \frac{1}{N_r} \sum_{r=1}^{N_r} \left\lVert C_r - \hat{C}_r \right\rVert_{1}
\end{equation}

\paragraph*{Masking Loss. }
Furthermore, to encourage the template scene represent only the object of interest, we use 2D segmentation masks to guide the optimization to the region where the object of interest lies. Specifically, we sum the weights along the ray and apply a binary cross-entropy loss between this sum and the corresponding mask value. In our experiments, we set $\lambda_{mask}$ as $0.1$ when optimizing for the template scene, and increase it to $1.0$ when tracking the template.

\begin{equation}
\mathcal{L}_{\mathrm{mask}} =
\frac{1}{N_r} \sum_{r=1}^{N_r} 
BCE(W_r, M_r)
\end{equation}

Where $W_r$ is the weights sum of samples on ray $r$ and $M_r$ is the mask value.

\paragraph*{Eikonal Loss. }
Additionally, to encourage the SDF to converge to a plausible SDF, we apply the eikonal regularization \cite{gropp2020implicitgeometricregularizationlearning} on the normal vectors by penalizing the deviation of their L2-norm from one as follows:

\begin{equation}
    \mathcal{L}_{\mathrm{eik}} = 
\frac{1}{N_r N_s} \sum_{r=1}^{N_r} \sum_{i=1}^{N_s} 
\left( \lVert \nabla_x f_{\mathrm{sdf}}(x_{r,i}) \rVert_2 - 1 \right)^2
\end{equation}

$\lambda_{eik}$ is set to be $0.1$ while optimizing for the template and while tracking it.

\subsection{Template Reconstruction} \label{sub:method:template}

The optimization in \cref{sub:method:optimization} requires an initialization for the explicit template $\fsdf$ and appearance $f_{rgb}$. 
We use the state-of-the-art static reconstruction method NeuralAngelo~\cite{li2023neuralangelo}, which achieves state-of-the-art performance among Neural SDF-based methods, on the multi-view images of the first frame to get this initialization. 
While the reconstruction of NeuralAngelo is very accurate, it cannot include information from later frames which might lead to small inaccuracies in the template due to occlusions or ambiguities, but also topological inconsistencies due to larger movement in later frames.
Our refinement during optimization, see \cref{sub:method:optimization}, allows us to increase the fidelity, remove mistakes and adapt the topology of the template without relying on a perfect reconstruction from NeuralAngelo.
With this, we can achieve a high-quality reconstruction for each frame in only $25k$ steps, which amounts to $10\%$ of the optimization time for the template scene.

\begin{figure*}
    \centering
    \setlength{\tabcolsep}{1pt} %
    \renewcommand{\arraystretch}{1} %
    \newcommand{\wwidth}[0]{0.135}

    \begin{tabular}{ccccccc}
        GauSTAR \cite{zheng2025gaustar} & Tensor4D \cite{shao2023tensor4d} & HumanRF \cite{isik2023humanrf} & \textbf{MoAngelo (ours)} & Ground-truth \\
        \rotatebox{90}{Actor1} \includegraphics[width=\wwidth\textwidth]{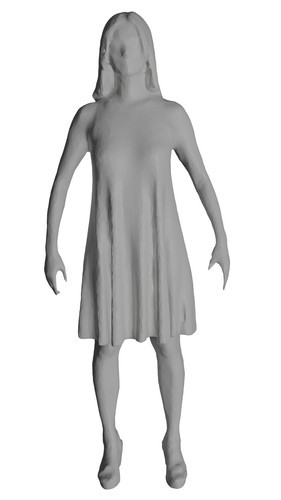} &
        \includegraphics[width=\wwidth\textwidth]{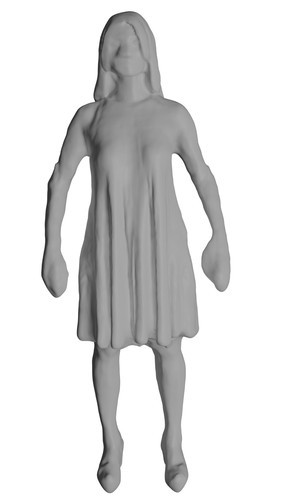} &
        \includegraphics[width=\wwidth\textwidth]{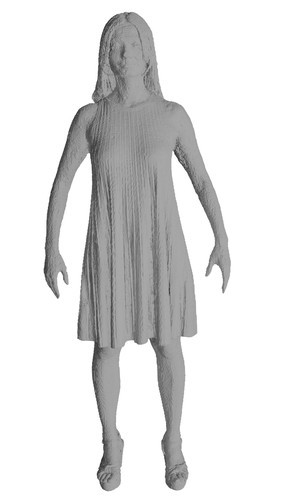} &
        \includegraphics[width=\wwidth\textwidth]{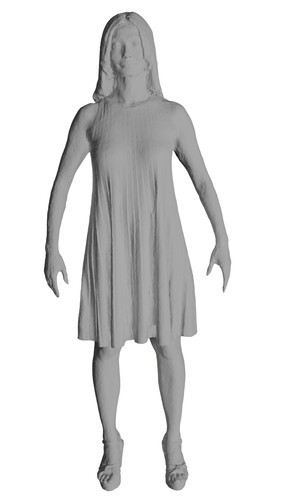} & 
        \includegraphics[width=\wwidth\textwidth]{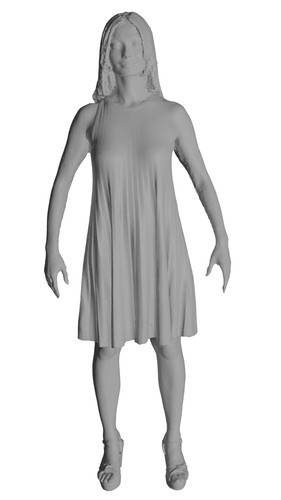} &

        \\
        \rotatebox{90}{Actor5} \includegraphics[width=\wwidth\textwidth]{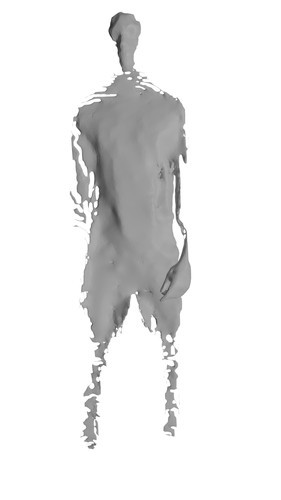} &
        \includegraphics[width=\wwidth\textwidth]{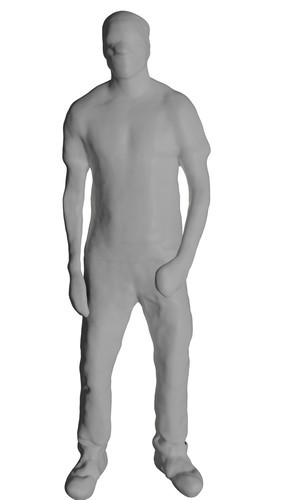} &
        \includegraphics[width=\wwidth\textwidth]{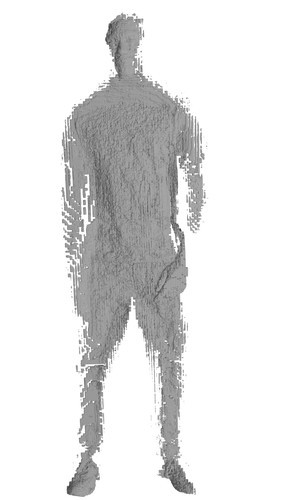} &
        \includegraphics[width=\wwidth\textwidth]{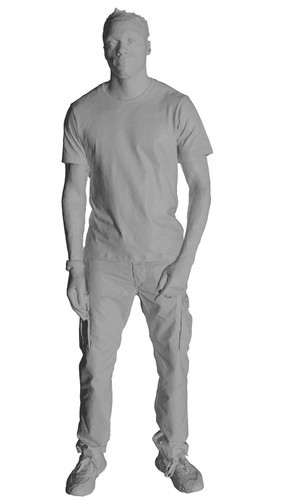} & 
        \includegraphics[width=\wwidth\textwidth]{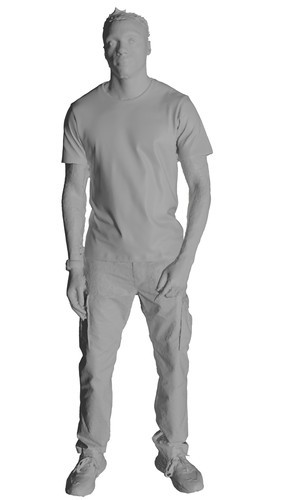} 
        \\
        \rotatebox{90}{Actor6} \includegraphics[width=\wwidth\textwidth]{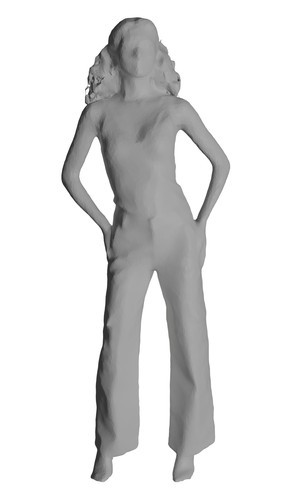} &
        \includegraphics[width=\wwidth\textwidth]{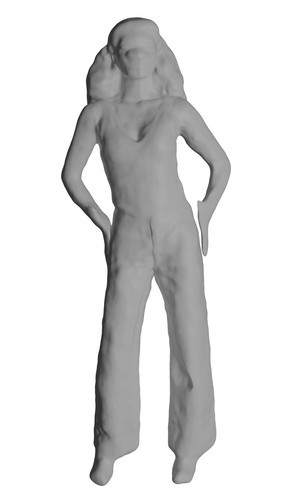} &
        \includegraphics[width=\wwidth\textwidth]{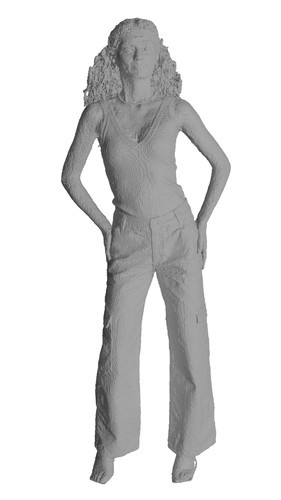} &
        \includegraphics[width=\wwidth\textwidth]{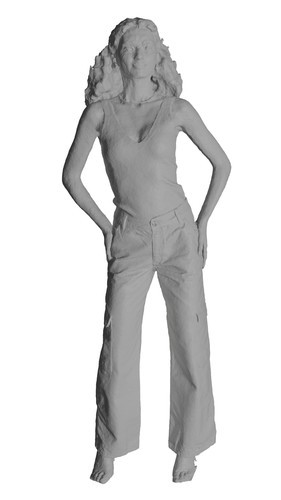} & 
        \includegraphics[width=\wwidth\textwidth]{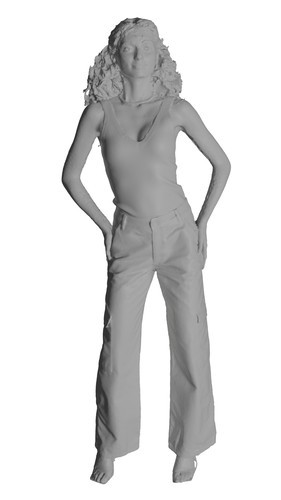} 
        \\
        \rotatebox{90}{Actor7} \includegraphics[width=\wwidth\textwidth]{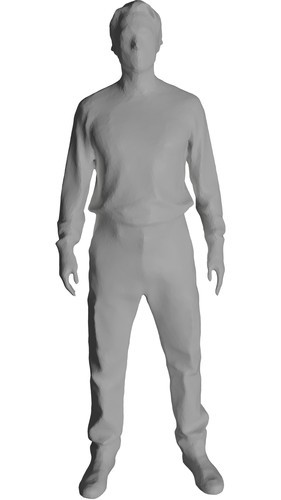} &
        \includegraphics[width=\wwidth\textwidth]{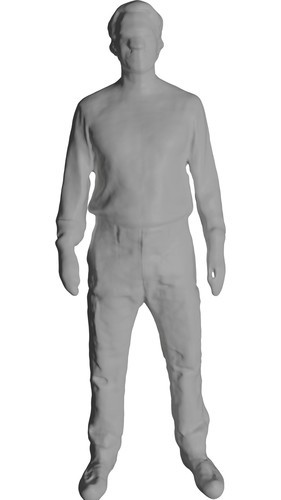} &
        \includegraphics[width=\wwidth\textwidth]{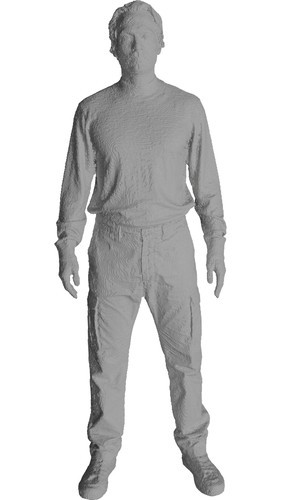} &
        \includegraphics[width=\wwidth\textwidth]{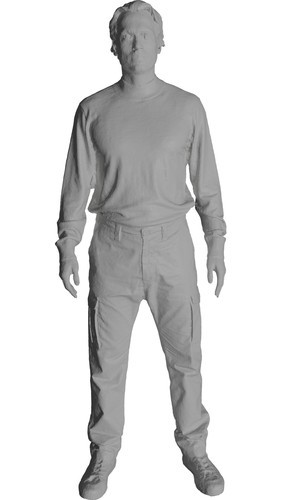} & 
        \includegraphics[width=\wwidth\textwidth]{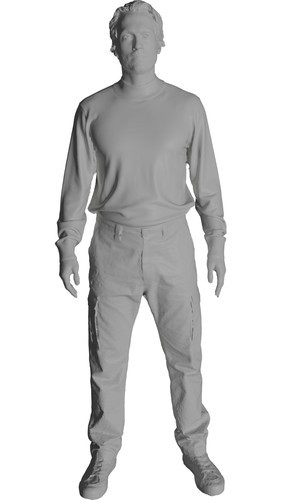} 
        \\
        \rotatebox{90}{Actor8} \includegraphics[width=\wwidth\textwidth]{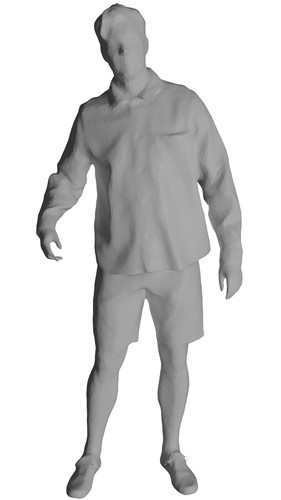} &
        \includegraphics[width=\wwidth\textwidth]{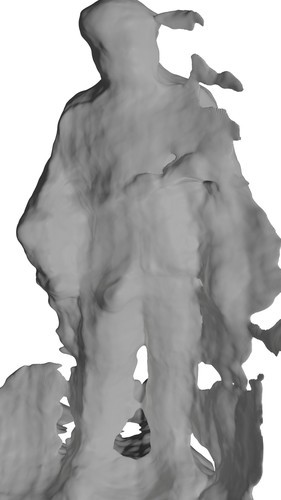} &
        \includegraphics[width=\wwidth\textwidth]{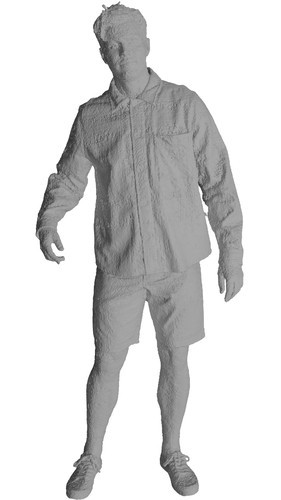} &
        \includegraphics[width=\wwidth\textwidth]{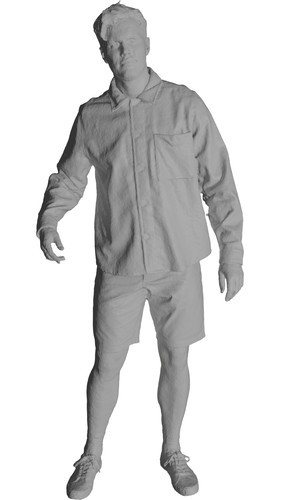} & 
        \includegraphics[width=\wwidth\textwidth]{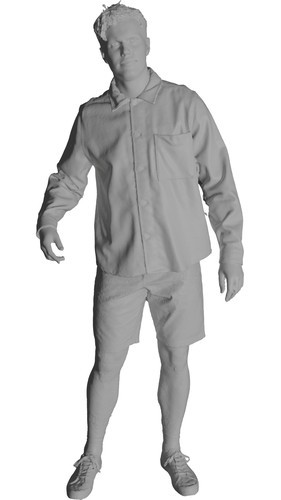} 
        \\

    \end{tabular}

    \caption{Qualitative results of competitors and MoAngelo (our method) from four different scenes in the ActorsHQ dataset. Our method reconstructs the highest fidelity geometry without unnecessary noise and does not have major failure cases.  }
    \label{fig:comparison}
\end{figure*}

\section{Evaluation} \label{sec:experiments}
\begin{figure*}
  \centering
  \newcommand{\wwidth}[0]{0.12}

  \begin{minipage}{0.9\textwidth}
    \makebox[\linewidth][c]{%
      \includegraphics[width=\wwidth\textwidth]{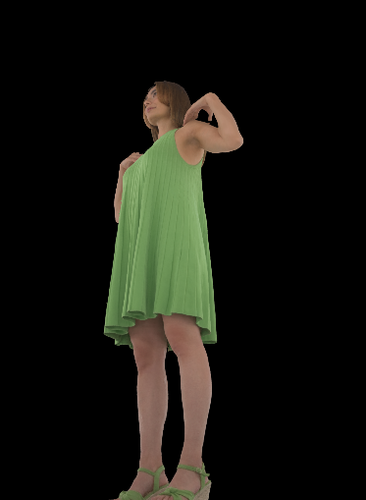}%
      \includegraphics[width=\wwidth\textwidth]{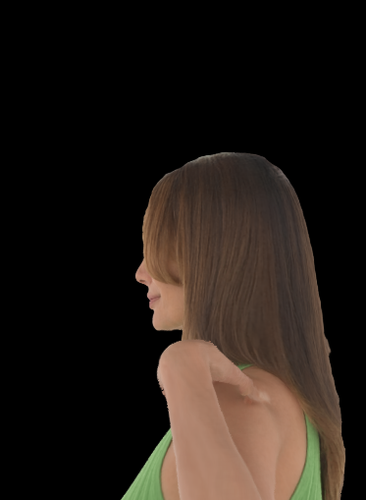}%
      \includegraphics[width=\wwidth\textwidth]{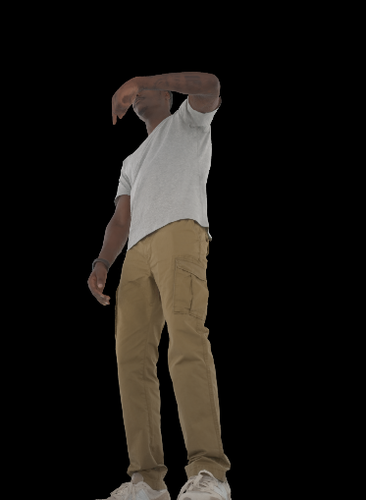}%
      \includegraphics[width=\wwidth\textwidth]{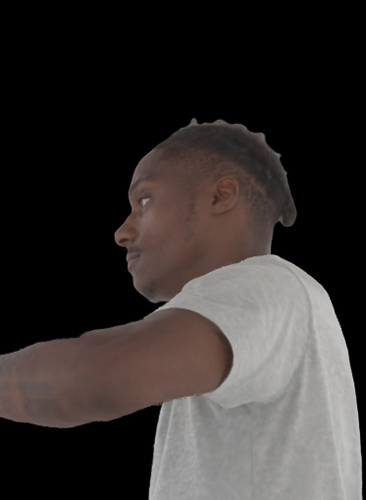}%
      \includegraphics[width=\wwidth\textwidth]{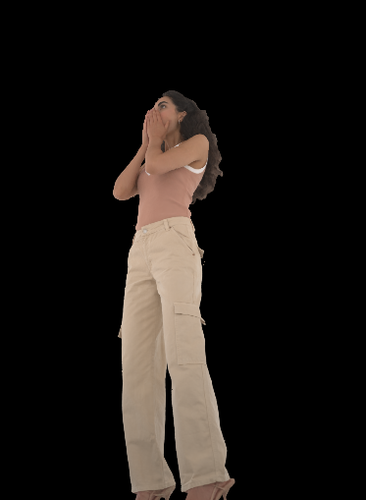}%
      \includegraphics[width=\wwidth\textwidth]{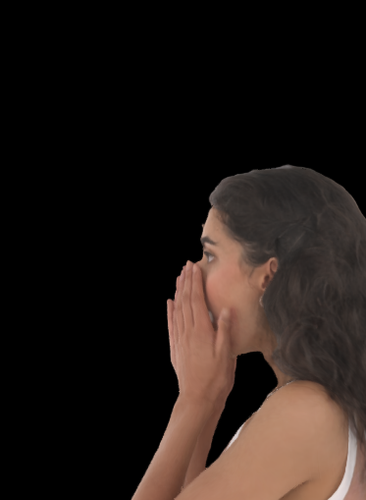}%
      \includegraphics[width=\wwidth\textwidth]{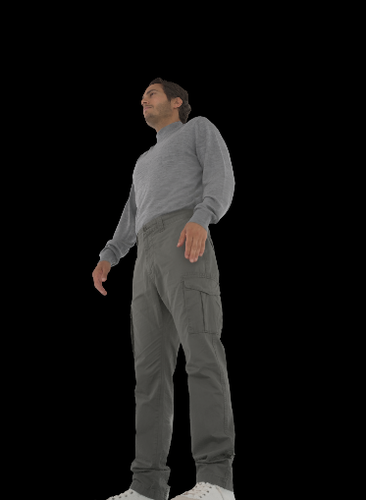}%
      \includegraphics[width=\wwidth\textwidth]{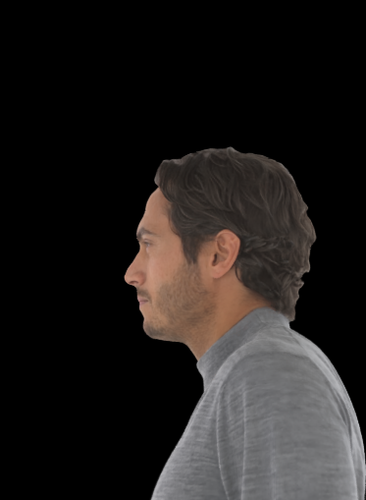}%
      \includegraphics[width=\wwidth\textwidth]{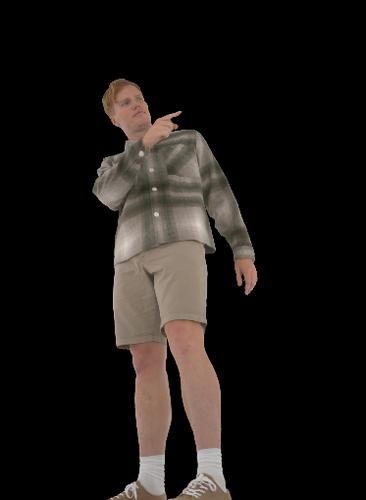}%
      \includegraphics[width=\wwidth\textwidth]{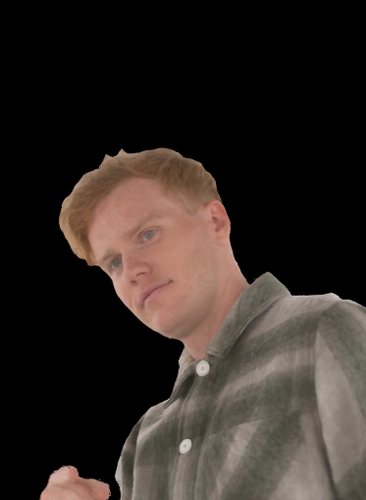}%
    }%
  \end{minipage}

  \begin{minipage}{0.9\textwidth}
    \makebox[\linewidth][c]{%
      \includegraphics[width=\wwidth\textwidth]{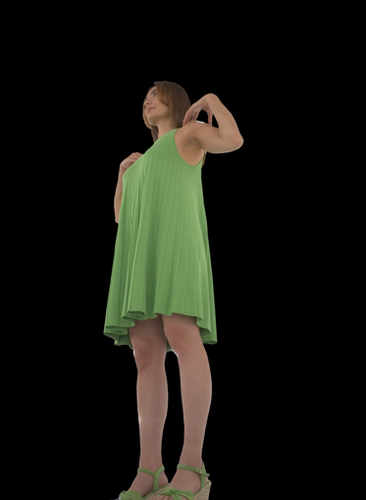}%
      \includegraphics[width=\wwidth\textwidth]{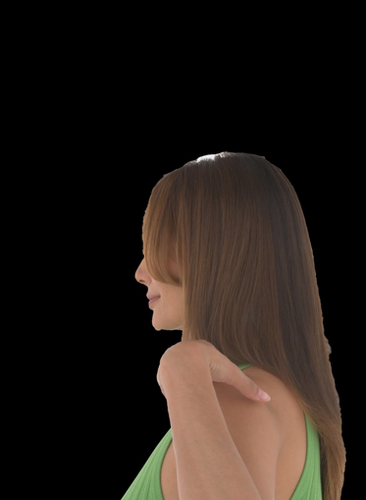}%
      \includegraphics[width=\wwidth\textwidth]{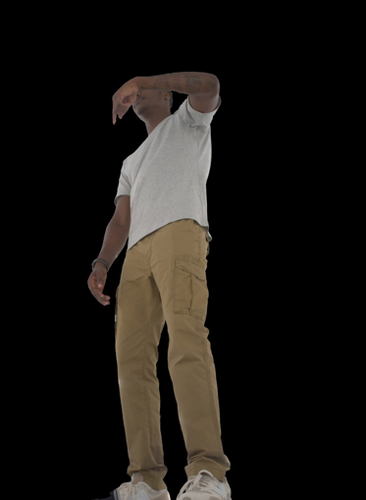}%
      \includegraphics[width=\wwidth\textwidth]{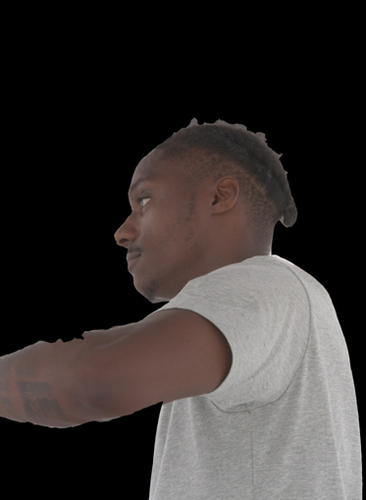}%
      \includegraphics[width=\wwidth\textwidth]{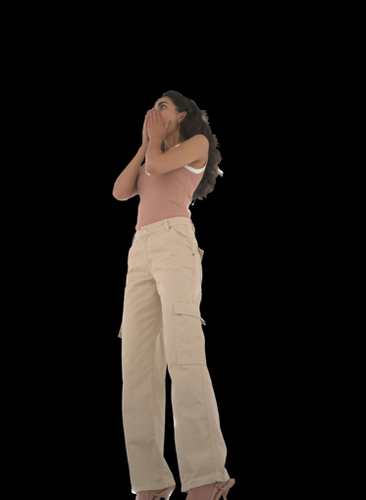}%
      \includegraphics[width=\wwidth\textwidth]{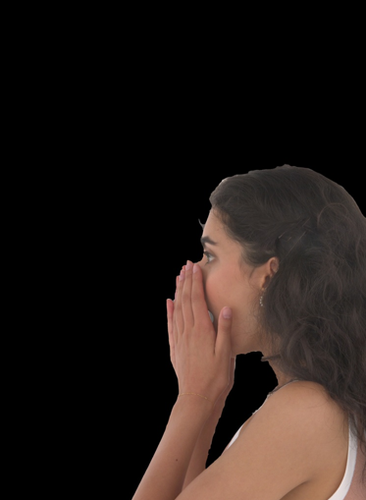}%
      \includegraphics[width=\wwidth\textwidth]{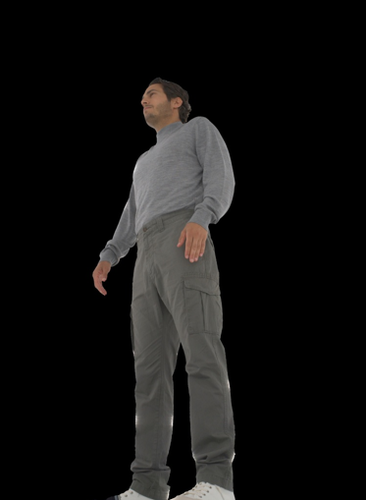}%
      \includegraphics[width=\wwidth\textwidth]{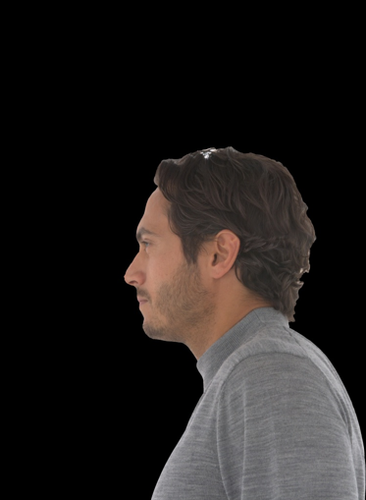}%
      \includegraphics[width=\wwidth\textwidth]{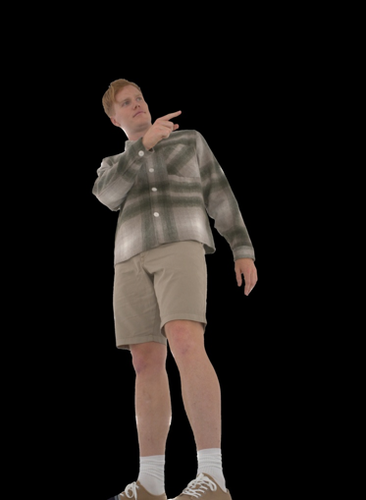}%
      \includegraphics[width=\wwidth\textwidth]{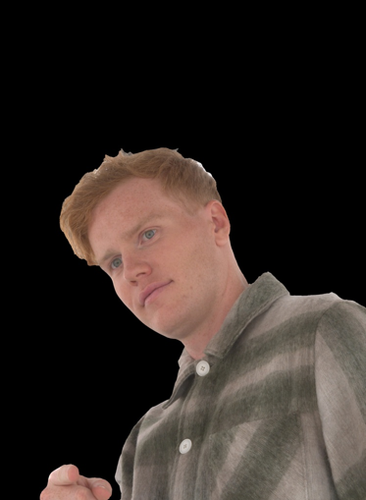}%
    }%
  \end{minipage}

  \caption{Visualization of renderings for novel-view synthesis. \textbf{Top:} Our renderings. \textbf{Bottom:} Ground truth, We demonstrate that our approach is capable not only of producing highly accurate meshes but also of generating high-quality renderings for novel-view synthesis
}
  \label{fig:ours-vs-gt}
\end{figure*}

\begin{figure}
    \centering
    \newcommand{\wwidth}{0.25} %

    \begin{minipage}{\wwidth\linewidth}
        \centering
        \includegraphics[width=\linewidth]{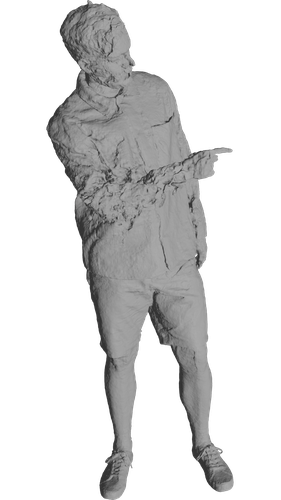}\\
        {\scriptsize w/o refinement}
    \end{minipage}\hfill
    \begin{minipage}{\wwidth\linewidth}
        \centering
        \includegraphics[width=\linewidth]{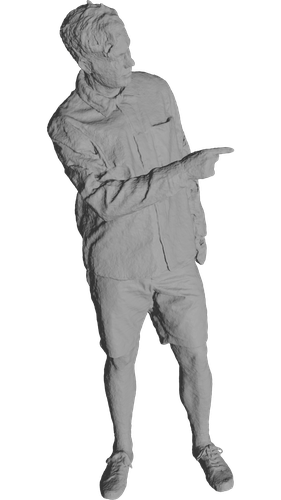}\\
        {\scriptsize w/o deform init}
    \end{minipage}\hfill
    \begin{minipage}{\wwidth\linewidth}
        \centering
        \includegraphics[width=\linewidth]{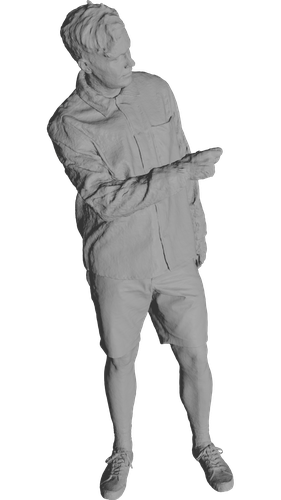}\\
        {\scriptsize w/o C2F deform}
    \end{minipage}\hfill
    \begin{minipage}{\wwidth\linewidth}
        \centering
        \includegraphics[width=\linewidth]{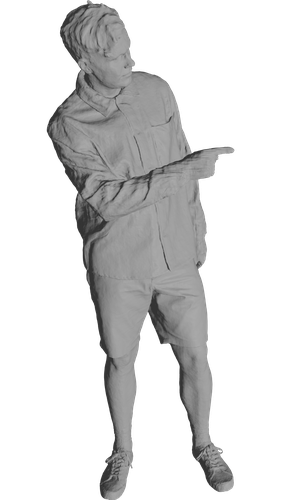}\\
        {\scriptsize \textbf{Ours (full)}}
    \end{minipage}

    \caption{Ablation study on the effect of different design choices. Our full model achieves the best reconstruction quality.}
    \label{fig:ablations}
\end{figure}

In the following section, we present the experimental evaluation of our proposed method against several baselines on dynamic reconstruction based on a quantitative evaluation~(\cref{sub:exp:quant}), qualitative results~(\cref{sub:exp:qual}), and an ablation study~(\cref{sub:exp:ablation}).
Additional qualitative examples can be found in the supplementary material. 
Experiments were done on a single Nvidia A100 GPU with 80GBs memory.

\subsection{Experimental Setup}

\subsubsection{Dataset}
To evaluate our method and compare it with the other baselines, we use the ActorsHQ dataset~\cite{isik2023humanrf}. It consists of multi-view videos of moving humans captured from 160 high-resolutional cameras with a total of 39,765 frames. 
We experiment on 6 scenes from the dataset with sequences where the actor is moving ranging from 30 to 40 time steps due to computational demands. 
We choose ActorsHQ because it the only multi-view video dataset that contains ground-truth meshes for all frames which we use to quantify and compare the reconstruction accuracy. 

\subsubsection{Baselines}
To show that our method achieves the best geometric quality for dynamic scenes, we compare with different recent dynamic approaches: \\
\textbf{Tensor4D \cite{shao2023tensor4d}} represents the geometry of the dynamic scene using a neural SDF defined by a 3D feature grid followed by a shallow MLP. They implicitly optimize for the template scene by having a deformation field aligning the positions of the observation frame to the coordinates of the template scene. The 3D and 4D hashgrids are decomposed into low-level tensors using tensor decomposition methods~\cite{chen2022tensorftensorialradiancefields}. \\
\textbf{HumanRF \cite{isik2023humanrf}} uses a neural field representation that maps the 4D input $(x,y,z,t)$ to a volume density and an RGB value. 
The neural field consists of a 4D grid followed by a shallow MLP.  The 4D grid is decomposed into lower rank multi-resolution hash grids using tensor decomposition. Since the geometry is encoded in a volume density, the extracted 3D meshes are low quality because the exact surface intersection is unknown. \\
\textbf{GauSTAR \cite{zheng2025gaustar}} combines 3D Gaussians splats with a 3D mesh. The initial template mesh is initialized from HumanRF\cite{isik2023humanrf}, and while the 3D Gaussians are tracked through time, the template mesh is deformed accordingly. 

\subsection{Quantitative Evaluation} \label{sub:exp:quant}
\begin{table}[t]
    \centering
    \scriptsize
    \begin{tabular}{ccccc}
        \toprule
        Scene & Tensor4D \cite{shao2023tensor4d} & GauSTAR \cite{zheng2025gaustar} & HumanRF \cite{isik2023humanrf} & Ours \\
        \midrule
        Actor1   & 0.0150      & 0.00833  & 0.0091  & \textbf{0.0057} \\
        Actor5   & 0.0187      & 0.2200  & 0.1100  & \textbf{0.0037} \\
        Actor6\_1 & 0.0160      & 0.0077  & 0.0087  & \textbf{0.0068} \\
        Actor6\_2 & 0.1830      & 0.0078  & 0.0088  & \textbf{0.0058} \\
        Actor7   & 0.0110 & 0.0067  & 0.0046  & \textbf{0.0035} \\
        Actor8   & 0.3590 & 0.0068  & 0.0119  & \textbf{0.0059} \\
        \midrule
        Average  & 0.1005 & 0.04288  & 0.0255  & \textbf{0.0052} \\
        \bottomrule
    \end{tabular}
    \caption{\textbf{Reconstruction Accuracy.} We report $L_1$-Chamfer distance (lower is better) for all methods. Our method provides the most accurate reconstruction on all scenes. }

    \label{tab:chamfer}
\end{table}

To evaluate the quality of our reconstructions quantitatively and to compare them with the baselines, we compute the Chamfer distance~\cite{chamfer_distance} between the reconstructed meshes and the ground truth meshes by sampling a point cloud of 1M points from each surface and computing the $L_1$ Chamfer distance between them. 
Results are reported in \cref{tab:chamfer} and we outperform all competitors on all scenes. 
To extract meshes from HumanRF, we use the postprocessing proposed in GauSTAR \cite{zheng2025gaustar}. 
Specifically, we apply marching cubes on the density field and use the provided 3D occupancy grids computed from the segmentation masks in the dataset to zero the non-occupied regions in the volume density grid.
For this reason, it is not be possible to extract a mesh from HumanRF if segmentation masks are not available. 
Our method does not suffer this limitation since the zero level set of an SDF is always well defined.

\subsection{Qualitative Evaluation} \label{sub:exp:qual}

\cref{fig:comparison} presents a direct qualitative comparison which shows that our method reconstructs much finer details than competitors and even provides accurate results in complex cases. 
A video of the reconstruction of all sequences in comparison to the baselines on ActorsHQ can be found here: \url{https://youtu.be/Fgm1ht2O_G8}. Additional qualitative figures can be found in the supplementary.

\subsection{Ablation Study} \label{sub:exp:ablation}

\begin{table}[t]
    \scriptsize
    \centering
    \begin{tabular}{lcccc}
        \toprule
        Scene & w/o refine. & w/o deform init. & w/o C2F deform & \textbf{Ours (full)} \\
        \midrule
        Actor1   & 0.0973  & 0.0057 & 0.0074 & \textbf{0.0057} \\
        Actor5   & 0.0359  & 0.0042 & \textbf{0.0037} & \textbf{0.0037} \\
        Actor6\_1 & 0.0088  & 0.0073 & 0.0070 & \textbf{0.0068}\\
        Actor6\_2 & 0.0062  & 0.0068 & 0.0062 & \textbf{0.0058} \\
        Actor7   & 0.0043  & 0.0039 & \textbf{0.0035} & \textbf{0.0035} \\
        Actor8   & 0.0089  & 0.0071 & 0.0080 & \textbf{0.0059} \\
        \midrule
        \textbf{Average} & 0.0269 & 0.0058 & 0.0059 & \textbf{0.0052}\\
        \bottomrule
    \end{tabular}
    \caption{\textbf{Ablation Study.} We report $L_1$-Chamfer distance (lower is better) for all the variants of our framework, our full version achieves the best chamfer distance on most scenes and on average}
    \label{tab:ablation}
\end{table}

In this section, we evaluate the different design choices of our framework, we run a quantitative comparison between the different variants and our full version using the same experimental settings mentioned before. We show in \cref{tab:ablation}, that our full version achieves the best average $L_1$-Chamfer distance on the scenes.

\subsubsection{Refining the Template Scene}
In our framework, gradients are propagated back to the template scene, allowing it to adapt to topology changes and occlusions that occur after the first frame. When this gradient flow is disabled and only the deformation field is optimized, the quality of the reconstruction is significantly degraded as shown in \cref{fig:ablations}. This degradation happens because the deformation field alone cannot capture topology changes or occlusions, as it can only model non-rigid transformations from the current frame to the template’s coordinate space.
\subsubsection{Initializing the Deformation Field}
In our experiments, we show that it is necessary to initialize the deformation field $f^{t}_{deform}$ with the weights of the previous deformation field $f^{t-1}_{deform}$, the experiments show that initializing the deformation field every time step from scratch makes it challenging for the deformation field to represent the correct deformation, the higher $t$, the larger is the transformation back to the template scene, leading to a noisy reconstruction as demonstrated in \cref{fig:ablations}.
\subsubsection{Coarse-To-Fine Optimization of the Deformation Field}
In our setup, we optimize the deformation field in a coarse-to-fine manner, we initially start with 4 active hashgrid levels, and gradually activate a new level every $1000$ steps. Initiating the optimization with the complete 16 hashgrid levels, lead to overfitting and the inability to accurately represent the deformation as shown in \cref{fig:ablations}.

\section{Conclusion}

We presented MoAngelo, a multi-view dynamic reconstruction method, based on deforming a template from a static reconstruction method to later frames while updating and refining the template during optimization. 
This joint update of geometry and deformation prevents oversmoothing, can update previously noisy or occluded parts and is flexible enough to incorporate later topology changes. 
In experiments, our method outperforms the previous state-of-the-art in reconstruction accuracy and provides significantly more visually pleasing results. 

{
    \small
    \bibliographystyle{ieeenat_fullname}
    \bibliography{main}
}
\setcounter{page}{1}
\maketitlesupplementary

\section{Additional Qualitative Results}
\begin{figure}
  \centering
  \newcommand{\wwidth}[0]{0.3}

  \begin{minipage}{0.9\linewidth}
    \makebox[\linewidth][c]{%
      \includegraphics[width=\wwidth\textwidth]{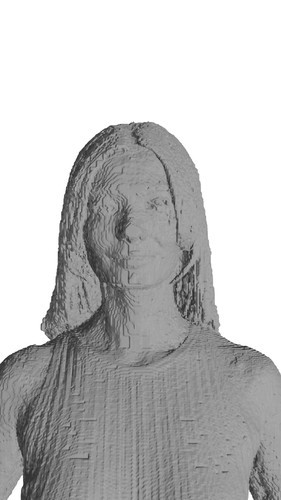}%
      \includegraphics[width=\wwidth\textwidth]{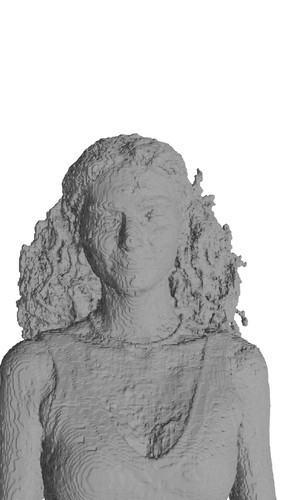}%
      \includegraphics[width=\wwidth\textwidth]{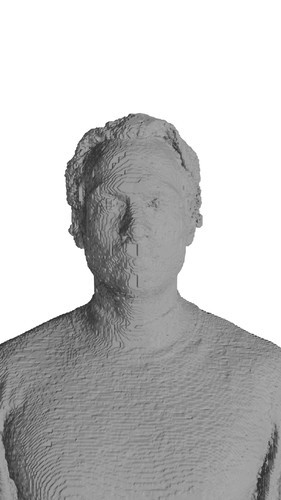}%
      \includegraphics[width=\wwidth\textwidth]{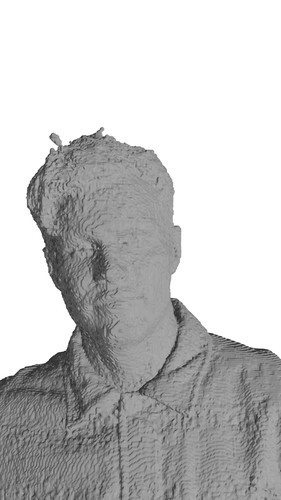}%
    }%
  \end{minipage}

  \begin{minipage}{0.9\linewidth}
    \makebox[\linewidth][c]{%
      \includegraphics[width=\wwidth\textwidth]{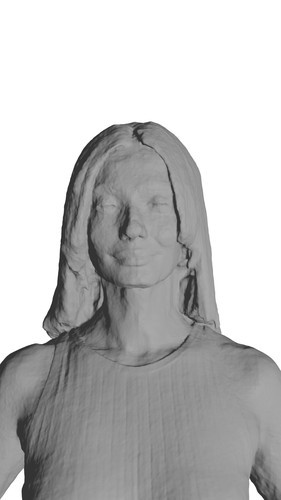}%
      \includegraphics[width=\wwidth\textwidth]{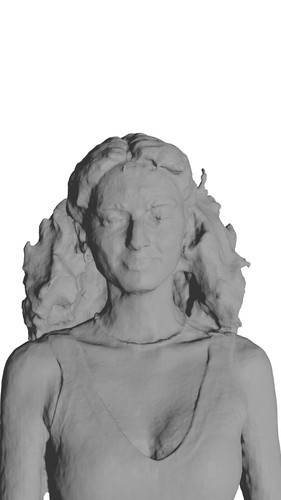}
      \includegraphics[width=\wwidth\textwidth]{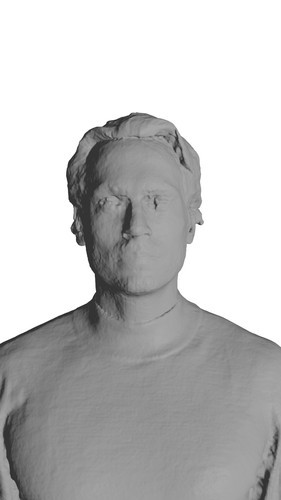}%
      \includegraphics[width=\wwidth\textwidth]{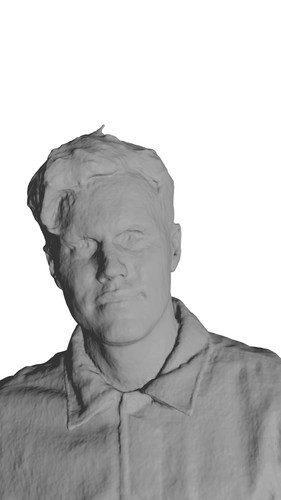}%
    }%
  \end{minipage}

  \caption{A close-up comparison between our method (bottom) and HumanRF~\cite{isik2023humanrf} (top). The results show that the reconstructions produced by HumanRF are noisy.
}
  \label{fig:ours-vs-humanrf}
\end{figure}

In this section, we present additional visualizations of our dynamic meshes across multiple frames, comparing them with the baselines. As illustrated in Figures~\ref{fig:comparison_actor1}, \ref{fig:comparison_actor5}, \ref{fig:comparison_actor6}, \ref{fig:comparison_actor7}, and \ref{fig:comparison_actor8}, our approach consistently surpasses the baselines in terms of reconstruction quality, particularly in recovering fine geometric details. For Actor5, we show in Figure~\ref{fig:comparison_actor5} a failure case where HumanRF~\cite{isik2023humanrf} and GauSTAR~\cite{zheng2025gaustar} fail. Since HumanRF represents geometry as a volume density, extracting an accurate surface is difficult, often resulting in an imprecise mesh, therefore GauSTAR also fails as the initial mesh is initialized from HumanRF.
We also provide a comparison in Figure~\ref{fig:ours-vs-humanrf}, which highlights that the reconstructions of HumanRF are noticeably noisier compared to ours. We additionally refer the reader to the supplementary video for a comparison between our dynamic meshes and the baselines.

\twocolumn[
{\begin{center}

    \rotatebox{90}{GauSTAR \cite{zheng2025gaustar}}
    \includegraphics[width=0.18\textwidth]{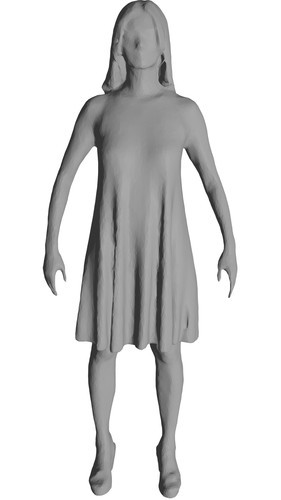}
    \includegraphics[width=0.18\textwidth]{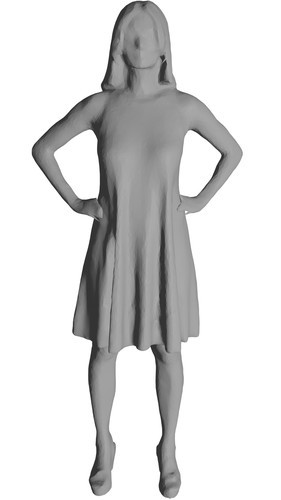}
    \includegraphics[width=0.18\textwidth]{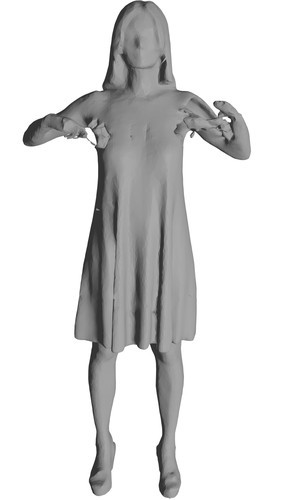}
    \includegraphics[width=0.18\textwidth]{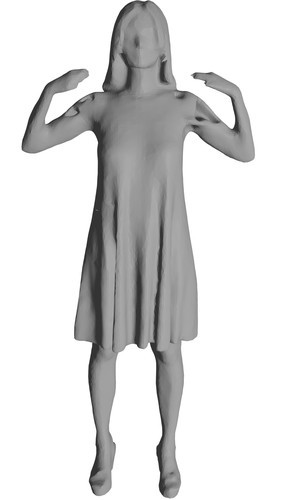}
    \includegraphics[width=0.18\textwidth]{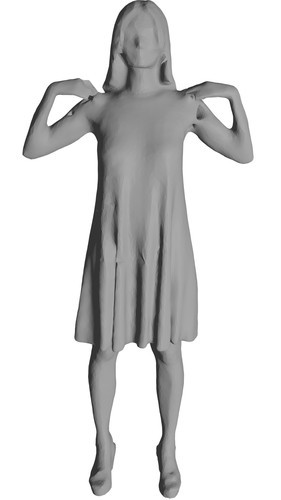}
    \\
    \rotatebox{90}{Tensor4D \cite{shao2023tensor4d}}
    \includegraphics[width=0.18\textwidth]{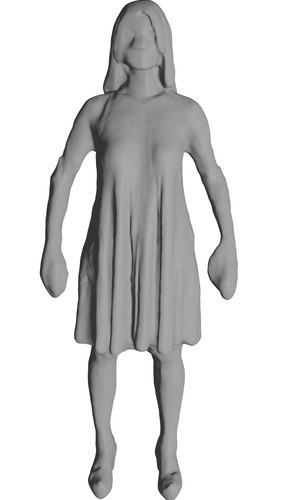}
    \includegraphics[width=0.18\textwidth]{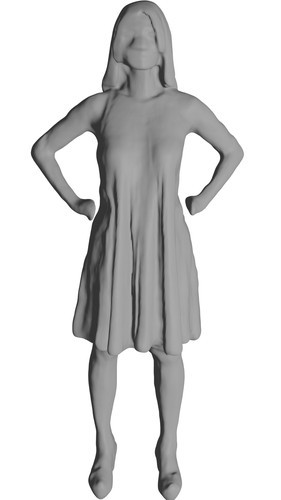}
    \includegraphics[width=0.18\textwidth]{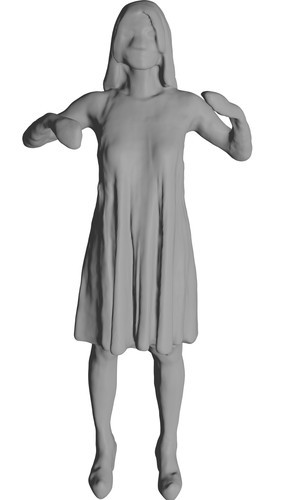}
    \includegraphics[width=0.18\textwidth]{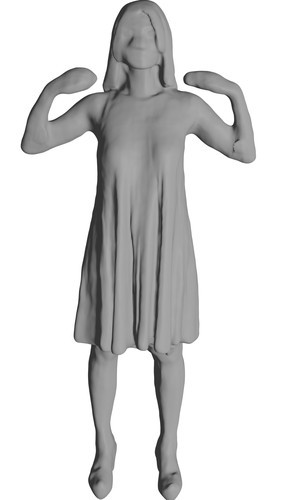}
    \includegraphics[width=0.18\textwidth]{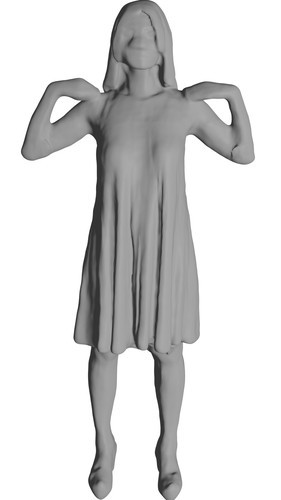}
    \\
    \rotatebox{90}{HumanRF \cite{isik2023humanrf}}
    \includegraphics[width=0.18\textwidth]{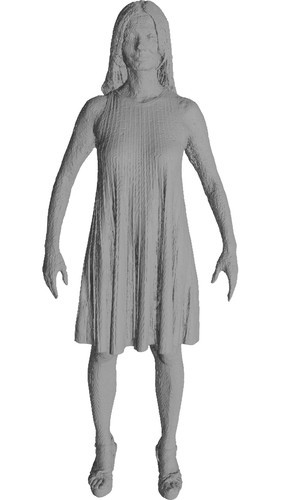}
    \includegraphics[width=0.18\textwidth]{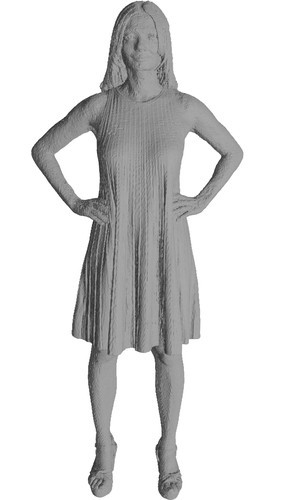}
    \includegraphics[width=0.18\textwidth]{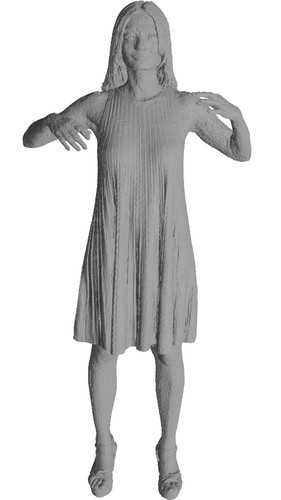}
    \includegraphics[width=0.18\textwidth]{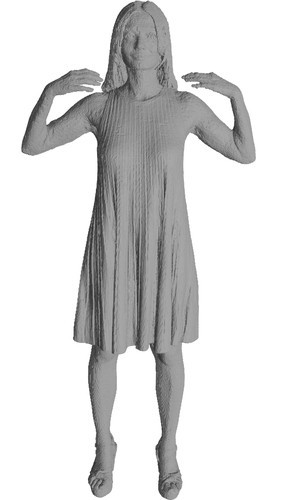}
    \includegraphics[width=0.18\textwidth]{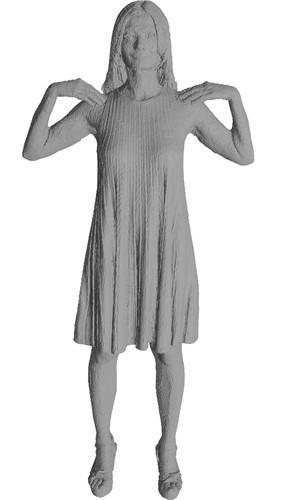}
    \\
    \rotatebox{90}{\textbf{Ours}}
    \includegraphics[width=0.18\textwidth]{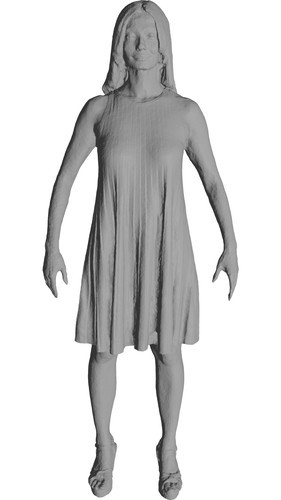}
    \includegraphics[width=0.18\textwidth]{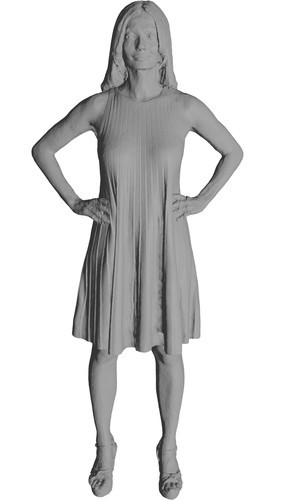}
    \includegraphics[width=0.18\textwidth]{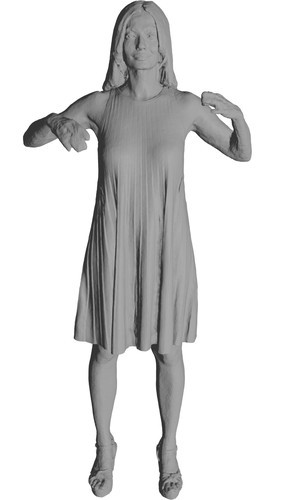}
    \includegraphics[width=0.18\textwidth]{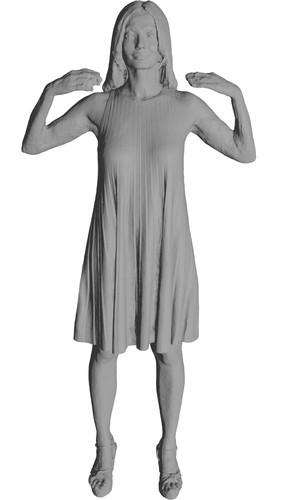}
    \includegraphics[width=0.18\textwidth]{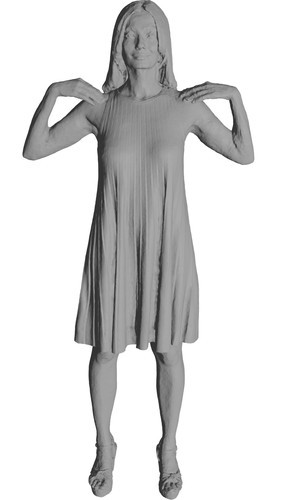}
    \\[-0.5em]

    \captionof{figure}{Additional Qualitative Comparison (Actor 1)}
    \label{fig:comparison_actor1}
\end{center}}
]

\twocolumn[
{\begin{center}

    \rotatebox{90}{GauSTAR \cite{zheng2025gaustar}}
    \includegraphics[width=0.18\textwidth]{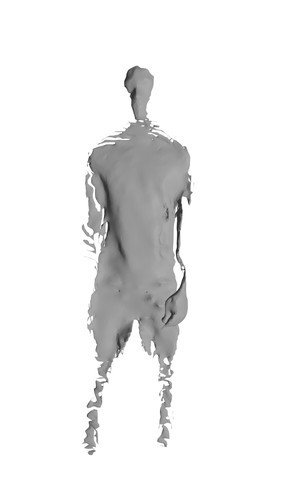}
    \includegraphics[width=0.18\textwidth]{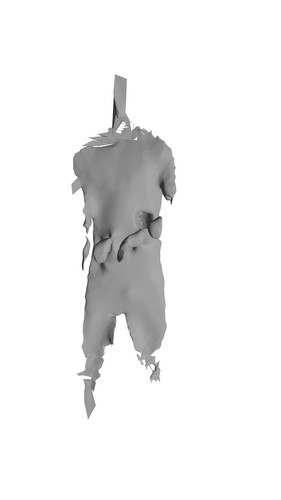}
    \includegraphics[width=0.18\textwidth]{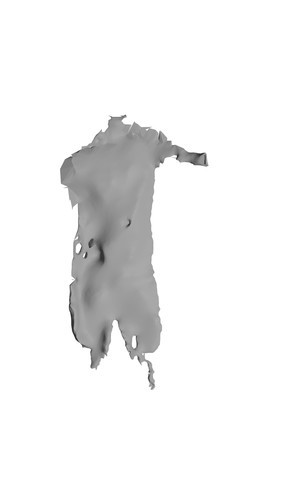}
    \includegraphics[width=0.18\textwidth]{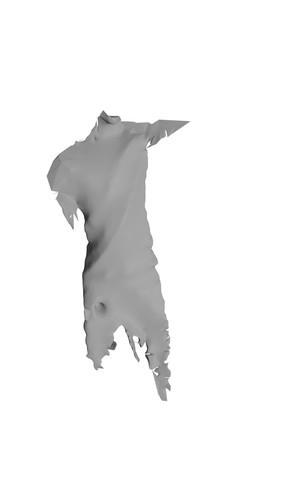}
    \includegraphics[width=0.18\textwidth]{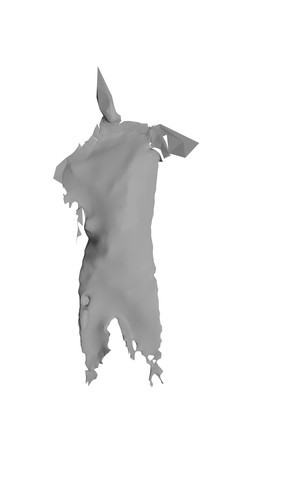}
    \\
    \rotatebox{90}{Tensor4D \cite{shao2023tensor4d}}
    \includegraphics[width=0.18\textwidth]{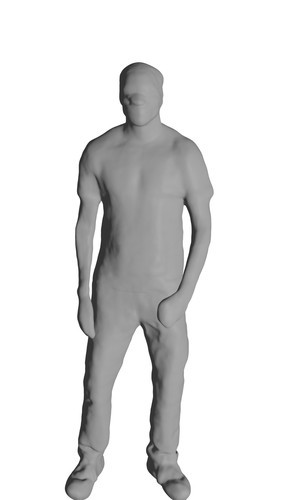}
    \includegraphics[width=0.18\textwidth]{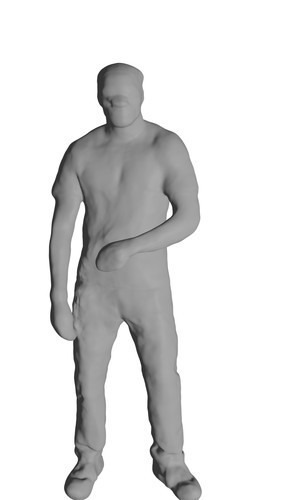}
    \includegraphics[width=0.18\textwidth]{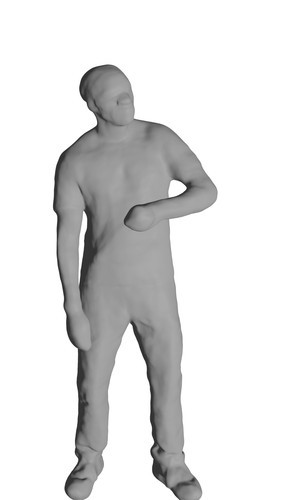}
    \includegraphics[width=0.18\textwidth]{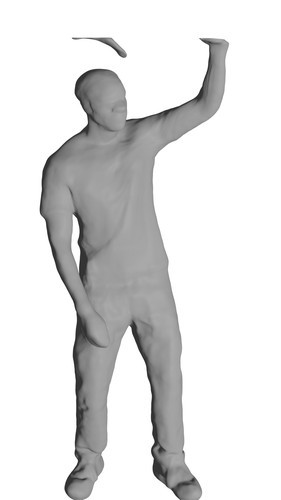}
    \includegraphics[width=0.18\textwidth]{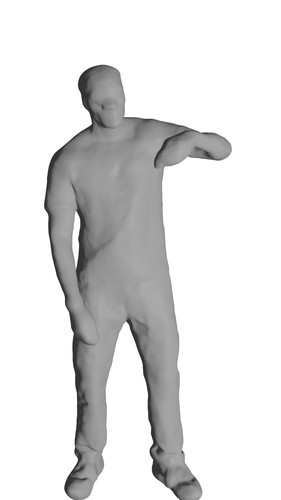}
    \\
    \rotatebox{90}{HumanRF \cite{isik2023humanrf}}
    \includegraphics[width=0.18\textwidth]{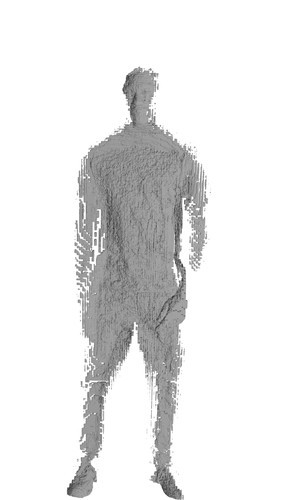}
    \includegraphics[width=0.18\textwidth]{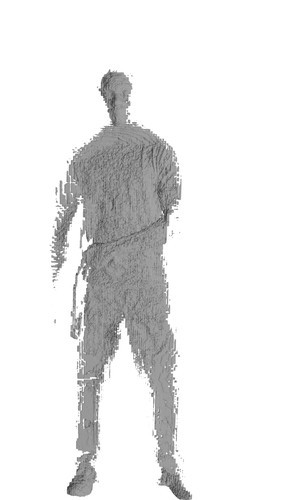}
    \includegraphics[width=0.18\textwidth]{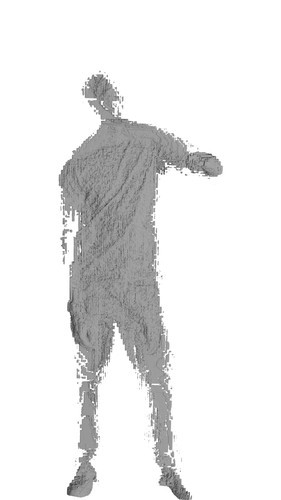}
    \includegraphics[width=0.18\textwidth]{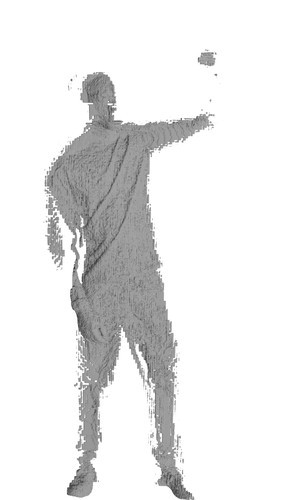}
    \includegraphics[width=0.18\textwidth]{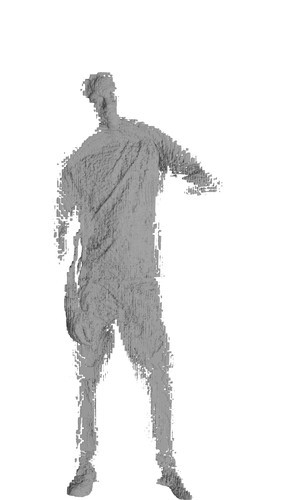}
    \\
    \rotatebox{90}{\textbf{Ours}}
    \includegraphics[width=0.18\textwidth]{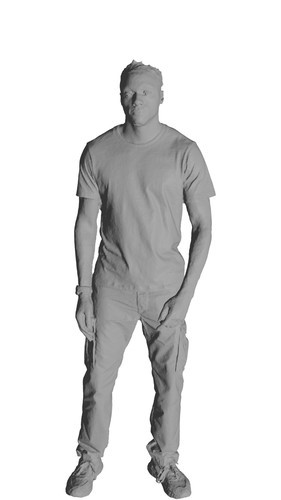}
    \includegraphics[width=0.18\textwidth]{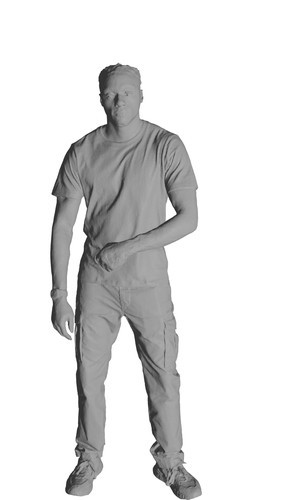}
    \includegraphics[width=0.18\textwidth]{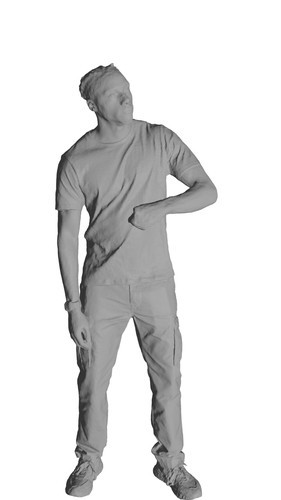}
    \includegraphics[width=0.18\textwidth]{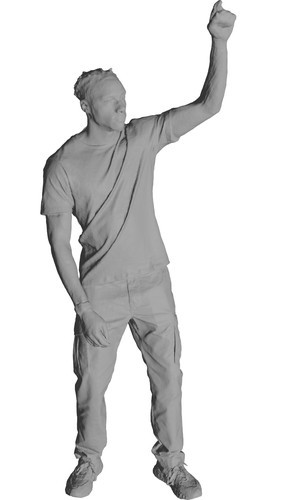}
    \includegraphics[width=0.18\textwidth]{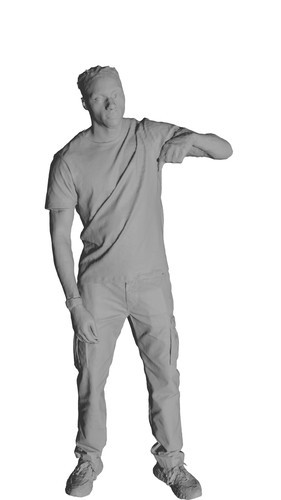}
    \\[-0.5em]

    \captionof{figure}{Additional Qualitative Comparison (Actor 5)}
    \label{fig:comparison_actor5}
\end{center}}
]

\twocolumn[
{\begin{center}

    \rotatebox{90}{GauSTAR \cite{zheng2025gaustar}}
    \includegraphics[width=0.18\textwidth]{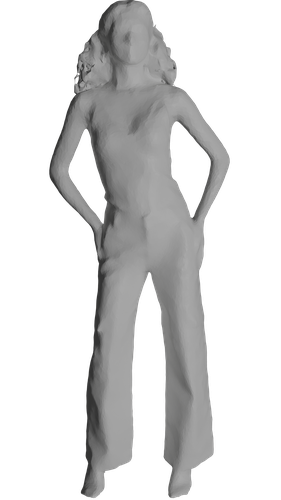}
    \includegraphics[width=0.18\textwidth]{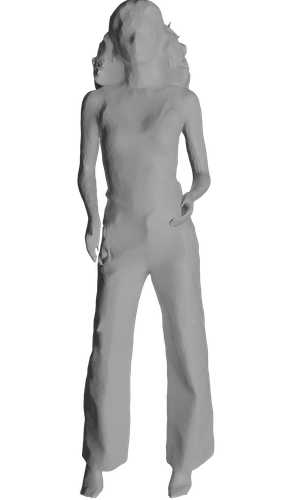}
    \includegraphics[width=0.18\textwidth]{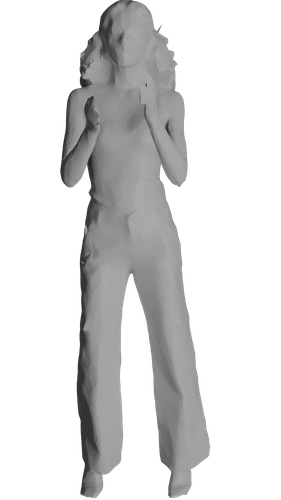}
    \includegraphics[width=0.18\textwidth]{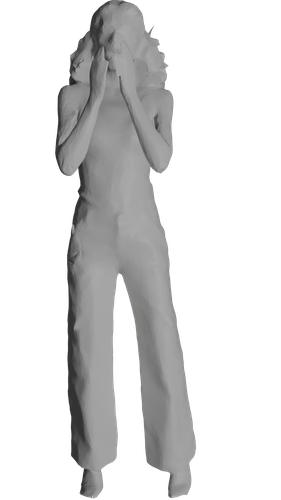}
    \includegraphics[width=0.18\textwidth]{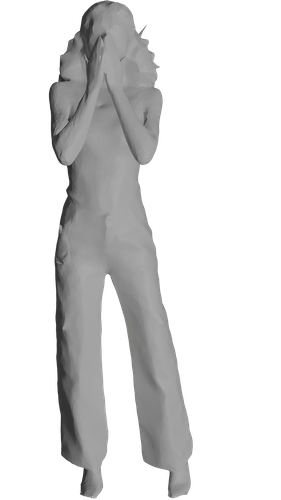}
    \\
    \rotatebox{90}{Tensor4D \cite{shao2023tensor4d}}
    \includegraphics[width=0.18\textwidth]{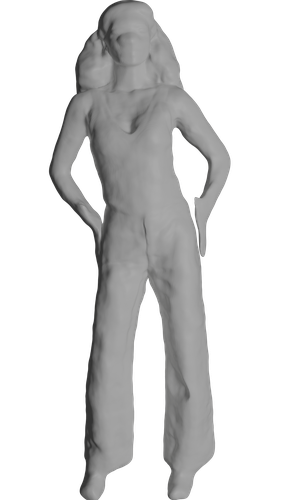}
    \includegraphics[width=0.18\textwidth]{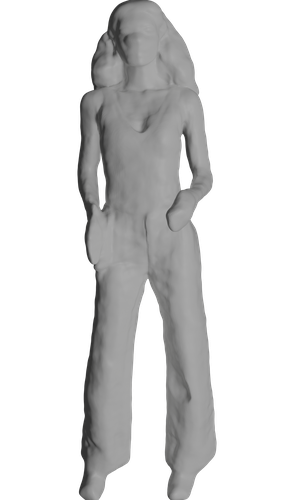}
    \includegraphics[width=0.18\textwidth]{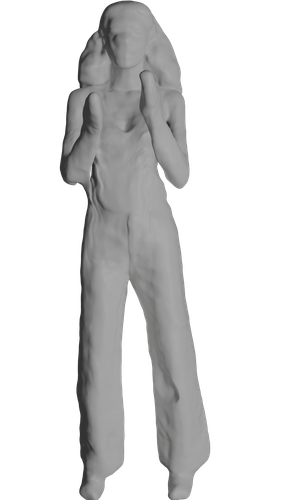}
    \includegraphics[width=0.18\textwidth]{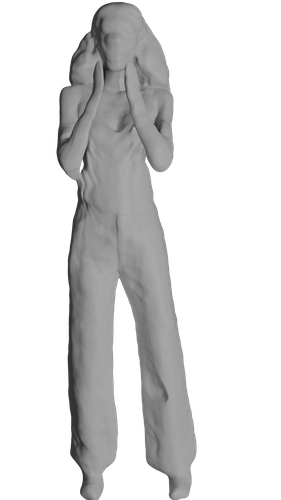}
    \includegraphics[width=0.18\textwidth]{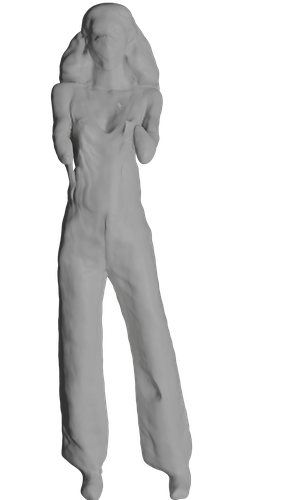}
    \\
    \rotatebox{90}{HumanRF \cite{isik2023humanrf}}
    \includegraphics[width=0.18\textwidth]{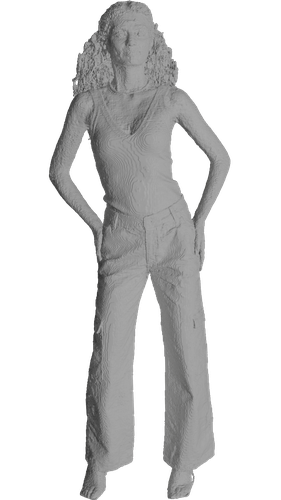}
    \includegraphics[width=0.18\textwidth]{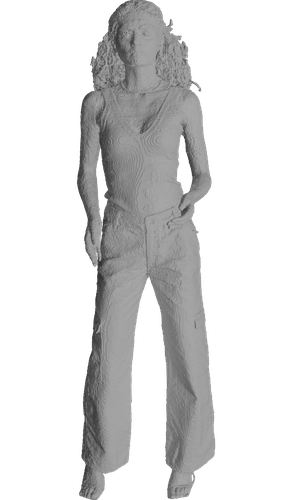}
    \includegraphics[width=0.18\textwidth]{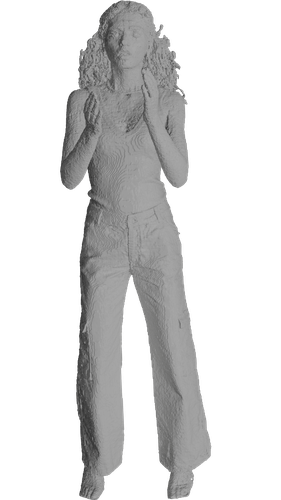}
    \includegraphics[width=0.18\textwidth]{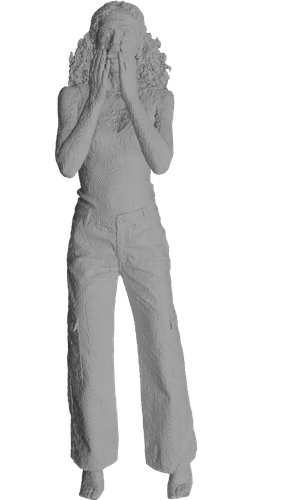}
    \includegraphics[width=0.18\textwidth]{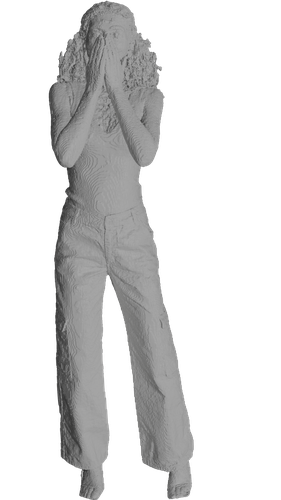}
    \\
    \rotatebox{90}{\textbf{Ours}}
    \includegraphics[width=0.18\textwidth]{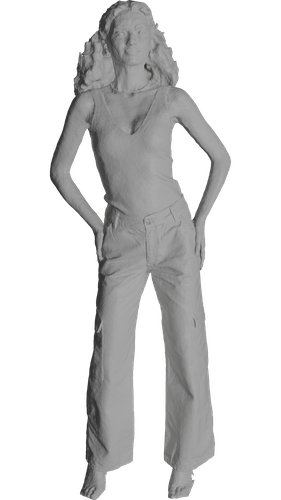}
    \includegraphics[width=0.18\textwidth]{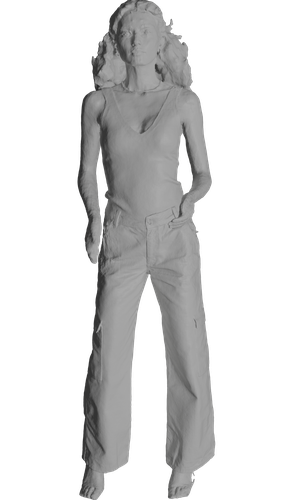}
    \includegraphics[width=0.18\textwidth]{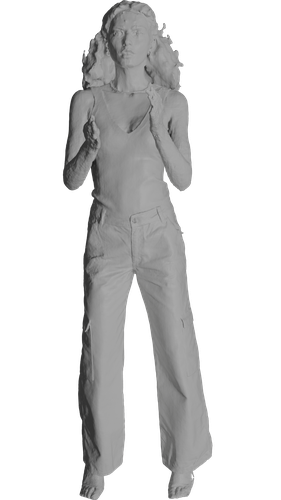}
    \includegraphics[width=0.18\textwidth]{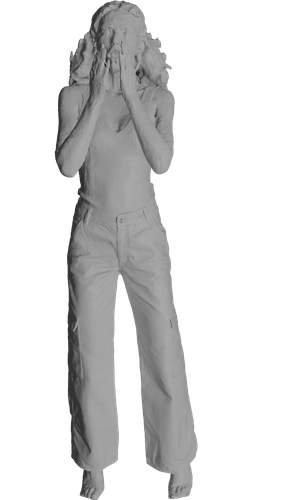}
    \includegraphics[width=0.18\textwidth]{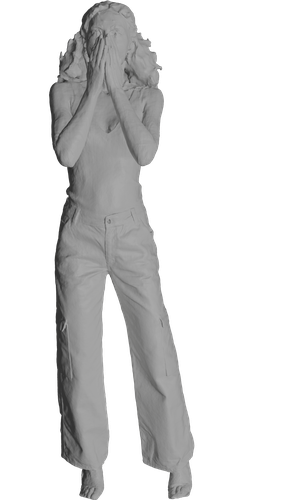}
    \\[-0.5em]

    \captionof{figure}{Additional Qualitative Comparison (Actor 6)}
    \label{fig:comparison_actor6}
\end{center}}
]

\twocolumn[
{\begin{center}

    \rotatebox{90}{GauSTAR \cite{zheng2025gaustar}}
    \includegraphics[width=0.18\textwidth]{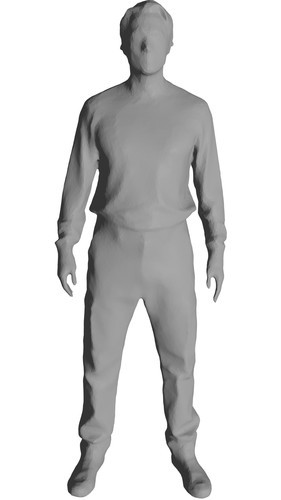}
    \includegraphics[width=0.18\textwidth]{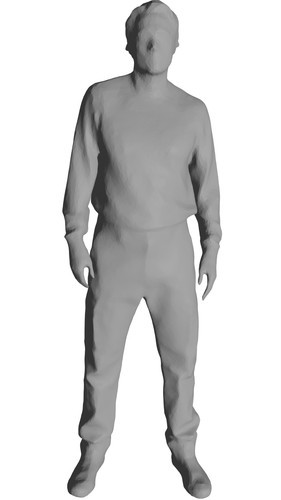}
    \includegraphics[width=0.18\textwidth]{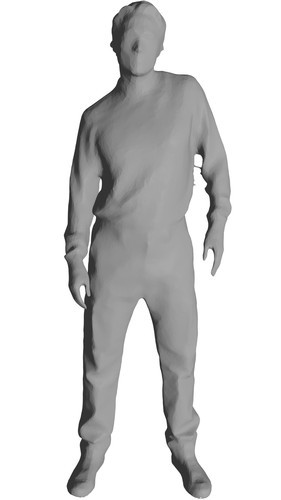}
    \includegraphics[width=0.18\textwidth]{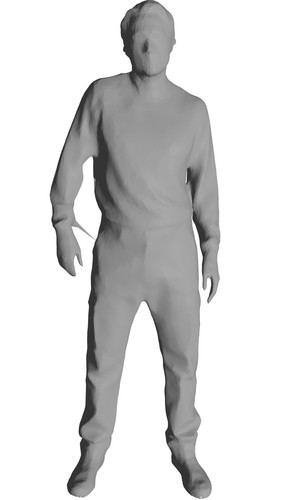}
    \includegraphics[width=0.18\textwidth]{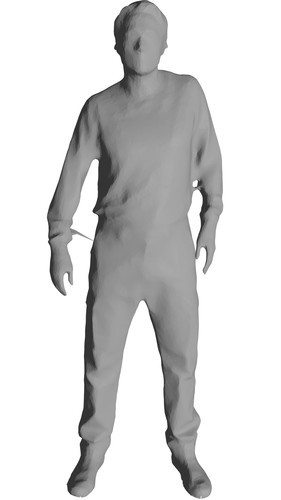}
    \\
    \rotatebox{90}{Tensor4D \cite{shao2023tensor4d}}
    \includegraphics[width=0.18\textwidth]{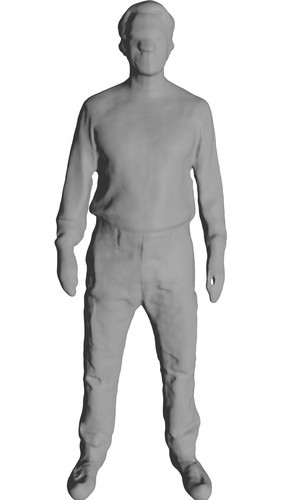}
    \includegraphics[width=0.18\textwidth]{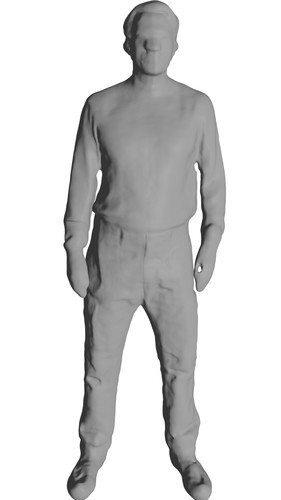}
    \includegraphics[width=0.18\textwidth]{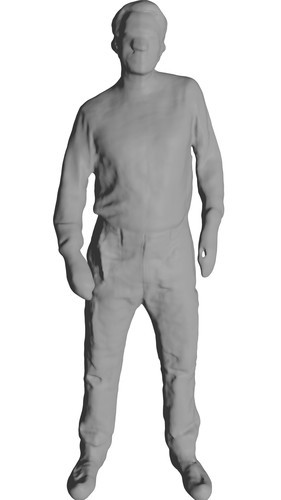}
    \includegraphics[width=0.18\textwidth]{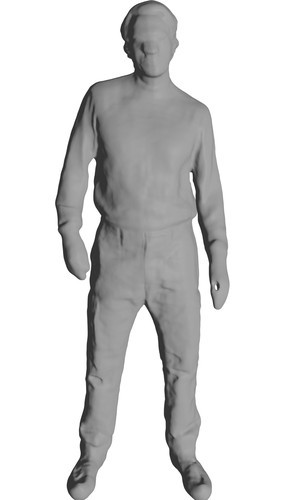}
    \includegraphics[width=0.18\textwidth]{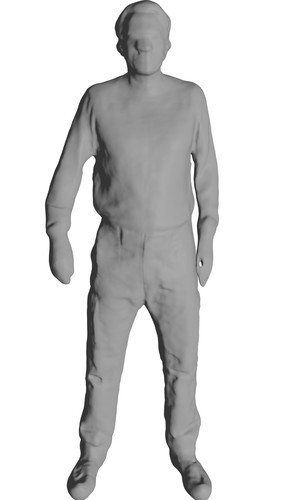}
    \\
    \rotatebox{90}{HumanRF \cite{isik2023humanrf}}
    \includegraphics[width=0.18\textwidth]{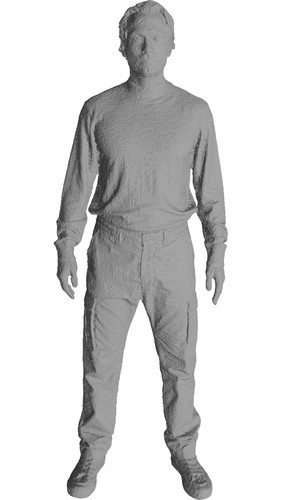}
    \includegraphics[width=0.18\textwidth]{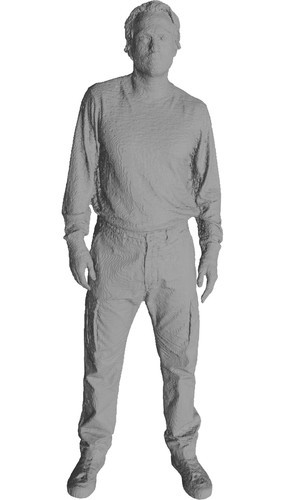}
    \includegraphics[width=0.18\textwidth]{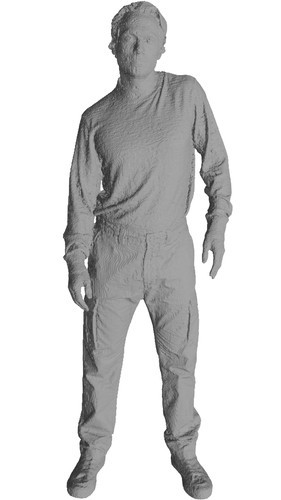}
    \includegraphics[width=0.18\textwidth]{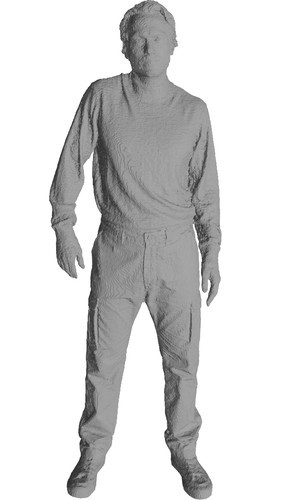}
    \includegraphics[width=0.18\textwidth]{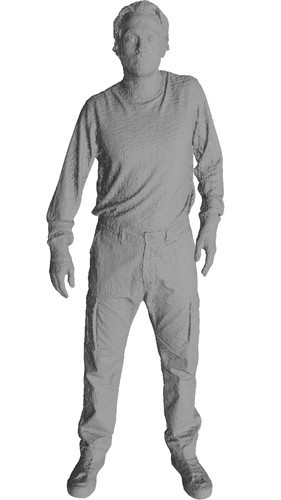}
    \\
    \rotatebox{90}{\textbf{Ours}}
    \includegraphics[width=0.18\textwidth]{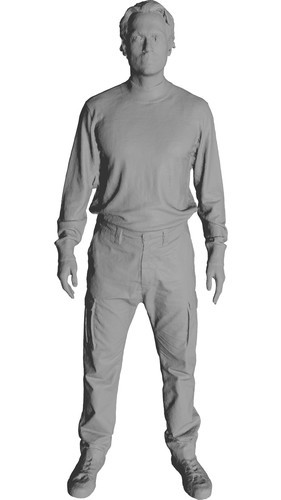}
    \includegraphics[width=0.18\textwidth]{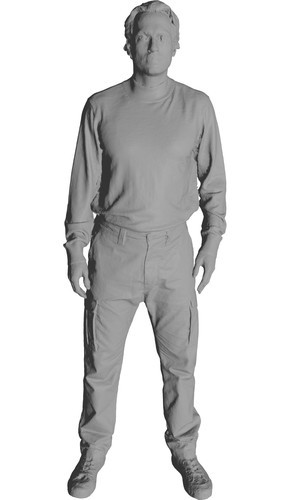}
    \includegraphics[width=0.18\textwidth]{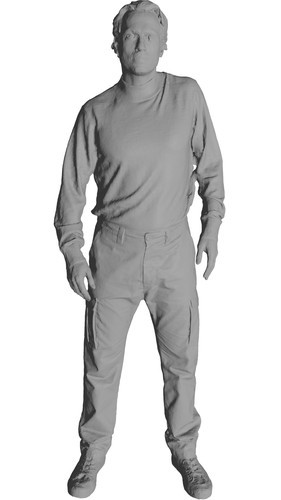}
    \includegraphics[width=0.18\textwidth]{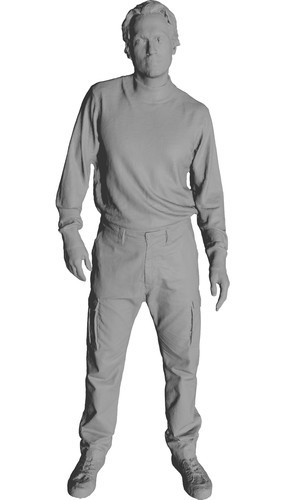}
    \includegraphics[width=0.18\textwidth]{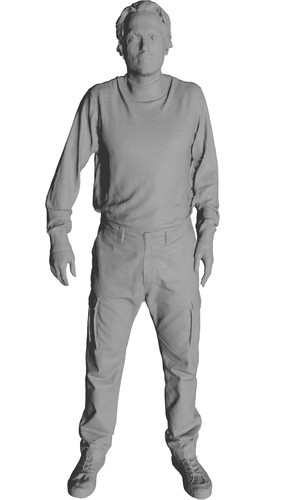}
    \\[-0.5em]

    \captionof{figure}{Additional Qualitative Comparison (Actor 7)}
    \label{fig:comparison_actor7}
\end{center}}
]

\twocolumn[
{\begin{center}

    \rotatebox{90}{GauSTAR \cite{zheng2025gaustar}}
    \includegraphics[width=0.18\textwidth]{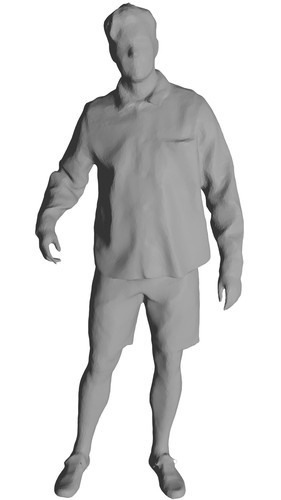}
    \includegraphics[width=0.18\textwidth]{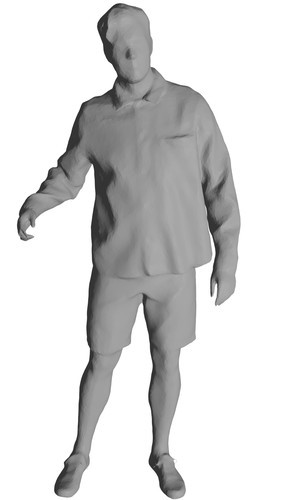}
    \includegraphics[width=0.18\textwidth]{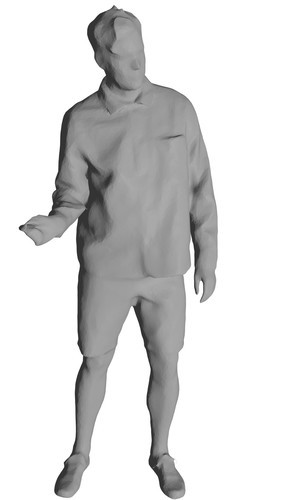}
    \includegraphics[width=0.18\textwidth]{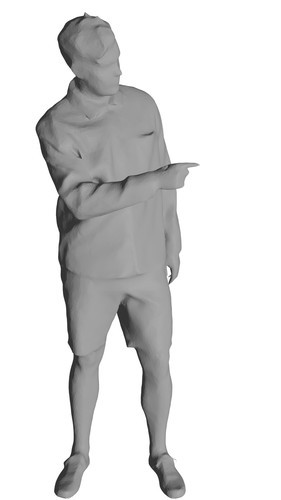}
    \includegraphics[width=0.18\textwidth]{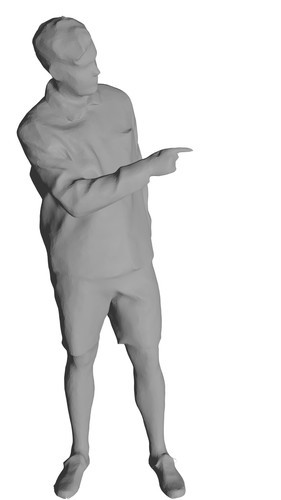}
    \\
    \rotatebox{90}{Tensor4D \cite{shao2023tensor4d}}
    \includegraphics[width=0.18\textwidth]{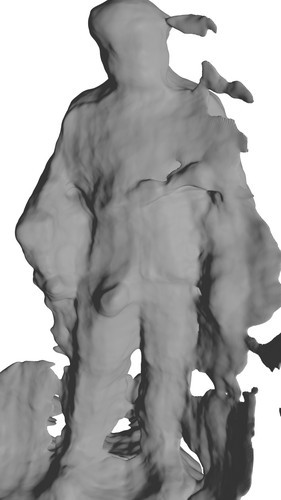}
    \includegraphics[width=0.18\textwidth]{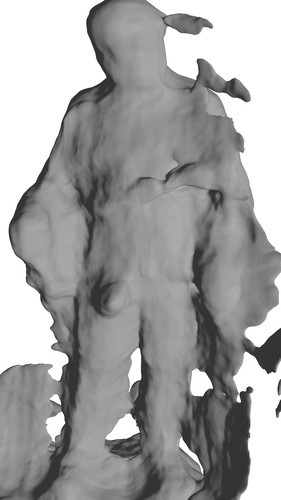}
    \includegraphics[width=0.18\textwidth]{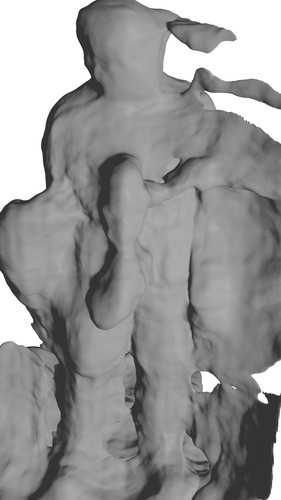}
    \includegraphics[width=0.18\textwidth]{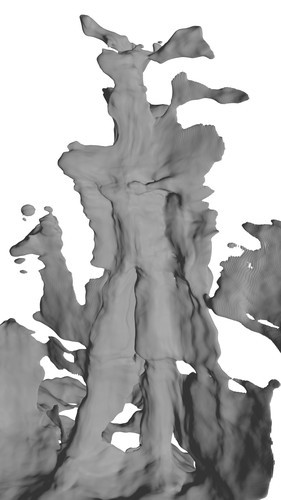}
    \includegraphics[width=0.18\textwidth]{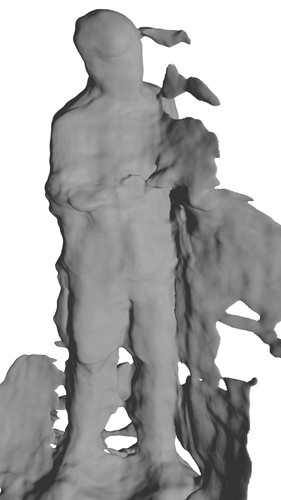}
    \\
    \rotatebox{90}{HumanRF \cite{isik2023humanrf}}
    \includegraphics[width=0.18\textwidth]{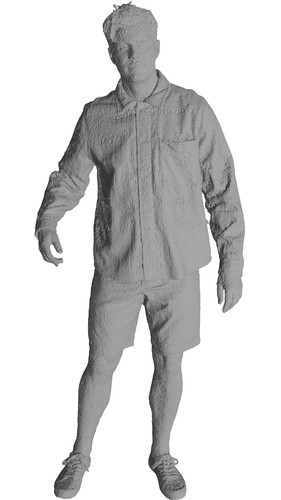}
    \includegraphics[width=0.18\textwidth]{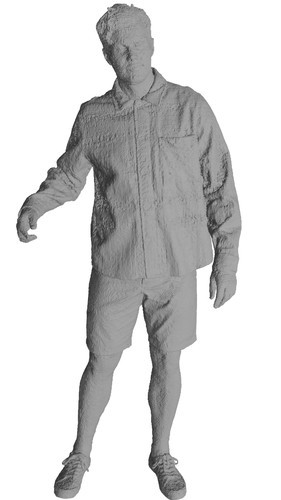}
    \includegraphics[width=0.18\textwidth]{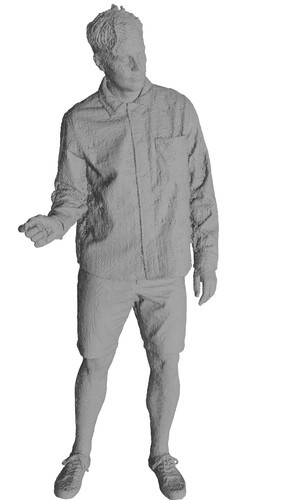}
    \includegraphics[width=0.18\textwidth]{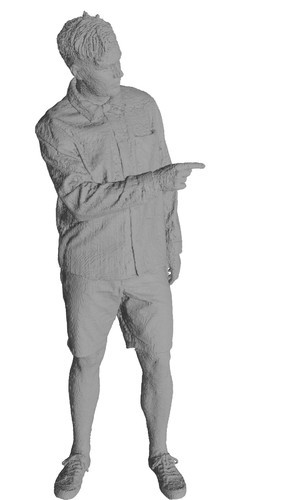}
    \includegraphics[width=0.18\textwidth]{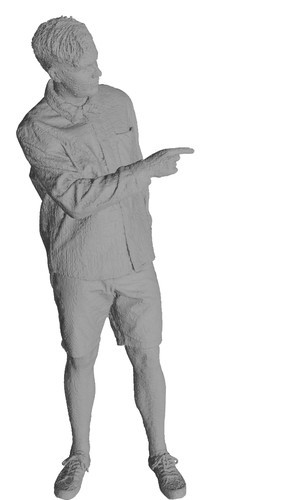}
    \\
    \rotatebox{90}{\textbf{Ours}}
    \includegraphics[width=0.18\textwidth]{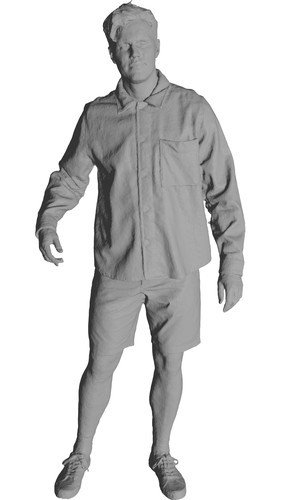}
    \includegraphics[width=0.18\textwidth]{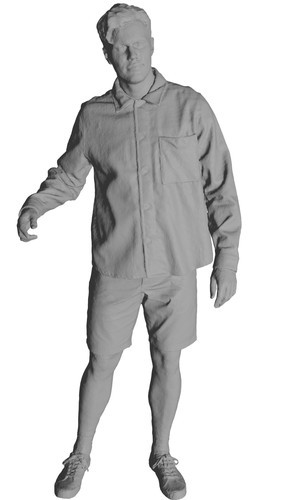}
    \includegraphics[width=0.18\textwidth]{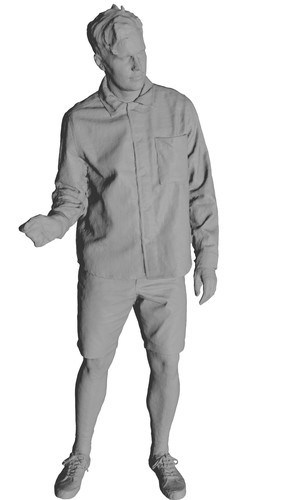}
    \includegraphics[width=0.18\textwidth]{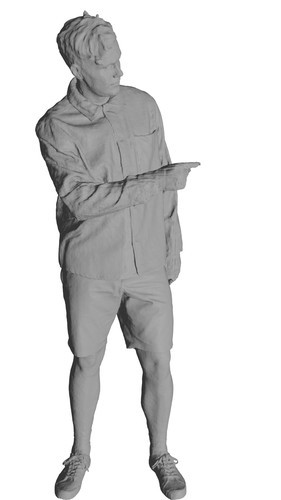}
    \includegraphics[width=0.18\textwidth]{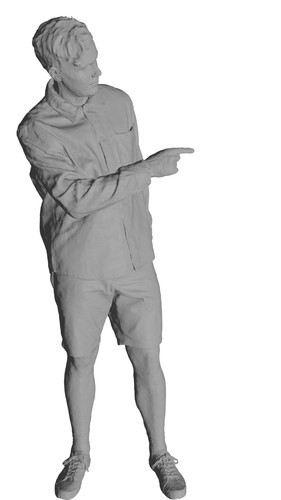}
    \\[-0.5em]

    \captionof{figure}{Additional Qualitative Comparison (Actor 8)}
    \label{fig:comparison_actor8}
\end{center}}
]

\end{document}